\documentclass[aps,showpacs]{revtex4}
\usepackage{graphicx}
\usepackage{float}
\usepackage[T1]{fontenc}
\usepackage[latin1]{inputenc}
\usepackage{amssymb}
\include{epsf}
\epsfverbosetrue

\newcommand{\xx}{\noindent}
\newcommand{\ra}{\rightarrow}
\newcommand{\Sign}{{\rm Sign}}
\newcommand{\Pslfrac}{P\!\!\!\!/~}
\newcommand{\Psl}{P\!\!\!\!/~}
\newcommand{\Ksl}{K\!\!\!\!/~}

\newcommand{\bea}{\begin{eqnarray}}
\newcommand{\eea}{\end{eqnarray}}

\makeatother
\begin{document}

\title{\Large{QED Electrical Conductivity using the 2PI Effective Action}}

\author{M.E. Carrington and E. Kovalchuk}

\email{carrington@brandonu.ca; kavalchuke@brandonu.ca}
 \affiliation{Department of Physics, Brandon University, Brandon, Manitoba, R7A 6A9 Canada\\ and \\
  Winnipeg Institute for Theoretical Physics, Winnipeg, Manitoba }

\begin{abstract}
In this article we calculate the electrical conductivity in QED using the 2PI effective action. We use a modified version of the usual 2PI effective action which is defined with respect to self-consistent solutions of the 2-point functions. We show that the green functions obtained from this modified effective action satisfy ward identities and that the conductivity obtained from the kubo relation is gauge invariant. We work to 3-loop order in the modified 2PI effective action and show explicitly that the resulting expression for the conductivity contains the square of the amplitude that corresponds to all binary collision and production processes. 
\end{abstract}

\maketitle
\section{Introduction}

Recent developments in heavy ion collisions and cosmology have stimulated interest in the theoretical understanding of the dynamics of quantum fields out of equilibrium. The 2PI effective action is one promising method to study such systems (for a review see \cite{berges}).

It is well known that calculations using the 2PI effective theory involve problems with gauge invariance. 
It is easy to see how these problems arise. In general, the ward identities depend on cancellations between different topologies that correspond to vertex corrections and self energy corrections. In a 2PI effective theory, one uses corrected propagators but not corrected vertices, and thus one does not expect the ward idenities to be satisfied. 

In this paper we study the applicability of the 2PI effective action to describe the equilibration of quantum fields. We consider the calculation of transport coefficients, which characterize the evolution of a system that is close to equilibrium over long time- and length-scales. We look specifically at the electrical conductivity of the QED plasma, which describes the diffusion of charge by an external electric field. Another important transport coefficient is the shear viscosity, which characterizes the diffusion of momentum transverse to the direction of propagation. The method we develop in this paper should be generalizable to the calculation of other transport coefficients.

The study of transport coefficients has a long history. In scalar theories they have been studied using finite temperature quantum field theory \cite{jeon1,hou1,heinz}, a direct ladder summation in Euclidean space \cite{basa}, and 2PI effective action methods \cite{gert1}. Gauge theories are more difficult to handle because of the subtlety of the power counting. The complete leading order calculation was done in \cite{AMY}. This calculation is not obtained directly from quantum field theory but is derived from kinetic theory.  The equivalence of the quantum field theory and  kinetic theory approaches has been demonstrated for scalar theories \cite{jeon1,hou1,heinz}. For gauge theories work has been done using the direct ladder summation in Euclidean space \cite{basa,gert2,hou2}, dynamical remormalization group methods \cite{rg}, and 2PI methods in the large $N_f$ approximation \cite{gert3}.  The conductivity has recently been calculated using a diagrammatic method in which the ward identity is used explicitly to select contributions that will produce a gauge invariant result \cite{jeon2}. 

In this paper we demonstrate how the calculation of transport coefficients is organized in the framework of the 2PI effective action. 
We show that a gauge invariant result can be obtained by constructing a new effective theory defined with respect to the self-consistent solutions for the 2-point functions. 
This type of strategy was originally proposed by Baym and Kadanoff \cite{baym} and has been discussed in the context of scalar theories in \cite{vanh}.  Vertices are obtained by taking functional deriatives of the modified effective action with respect to the expectation values of the fields and the self-consistent solutions for the 2-point functions. These vertices obey a set of bethe-salpeter type equations which effectively restore the crossing symmetry and allow one to obtain green functions that explicitly satisfy the ward identity. Equivalently, when using the modified effective action to calculate the conductivity, the summation over ladder graphs is obtained automatically, independent of any power counting analysis.

In addition to discussing the general properties of the modified 2PI effective action, we perform an explicit calculation at 3-loop order. We obtain the integral equation that determines the conductivity. As shown in \cite{gert3}, the 2-loop term produces the square of the $s$-channel which gives the complete result at the leading-log order of accuracy. We show that the 3-loop term produces the missing contributions to the $t$- and $u$- channels so that the full matrix element corresponding to all binary scattering and production processes is obtained. We note that this is not the complete leading order result, since the colinear terms are not included. These terms will be present in a calculation using the 3-loop 3PI effective action, and this work is currently in progress.

Our calculation provides a field theoretic connection to the kinetic theory results of \cite{AMY}, which is useful in itself. In addition, it seems likely that quantum field theory provides a better framework than kinetic theory for calculations beyond leading order. Our results provide strong support for the use of $n$PI effective theories as a method to study the equilibration of quantum fields. 

This paper is organized as follows. 

In section \ref{Notation} we define some notation. In \ref{CTP} we discuss the closed time path  formalism of real time statistical field theory which we use throughout this paper.  In \ref{PandV} we define the notation we use for propagators and vertices. 

In section \ref{2PI} we discuss the 2PI formalism. In \ref{basic2PI} we give the basic structure of the 2PI effective action for QED. In \ref{tildeGamma} we define the modified effective action as a function of the self-consistent solutions of the 2-point functions, and define external propagators and effective vertices. In  \ref{wi} we show that the photon propagator obeys the usual ward identity. In \ref{BetaSalEqn} we derive the bethe-salpeter type integral equations satisfied by the vertices.

In section \ref{Conductivity} we present the calculation of the integral equation that determines the conductivity. In \ref{kubo} we give the expression for the conductivity obtained from the kubo formula. In \ref{keldBS} we obtain a bethe-salpeter type integral equation for the 3-point vertex from the self-consistent constraint on the 2PI effective action. In \ref{ME} we discuss the structure of each contribution to this equation and show that the complete set of diagrams includes the scattering amplitudes for all binary scattering and production processes.  In \ref{IntEqn} we show that this integral equation is the same as the equation obtained in \cite{AMY} using kinetic theory. 

In section \ref{Conclusions} we present our conclusions and discuss future directions. 

In Appendix \ref{appendixA} we give some of the technical details of the calculations presented in section \ref{ME}.

\section{Notation}
\label{Notation}

\subsection{Keldysh Representation of Real Time Finite Temperature Field Theory}
\label{CTP}

Throughout this paper we use the closed time path formulation of real time statistical
field theory \cite{Sch,Keld} which consists of a contour with two
branches: one runs from minus infinity to infinity along the real
axis, the other runs back from infinity to minus infinity just below
the real axis (for reviews see, for example, \cite{gelis,MCTF}). The closed time path contour results in a doubling of degrees of
freedom.  Physically, these extra contributions come from the
additional processes that are present when the system interacts with a
medium, instead of sitting in a vacuum.  As a result of these extra
degrees of freedom, $n$-point functions have a tensor stucture. Statistical field theory
can be formulated in different bases, which produce different
representations of these tensors. We will work in the keldysh basis. 
In the rest of this section indices in the 1-2 basis will
be denoted $b_i$ and take the values 1 or 2.  Keldysh indices will
be written $c_i$ and are assigned the values $c_i = 1 := r$
and $c_i = 2:= a$.

Throughout this section we discuss scalar fields only. The keldysh structure of the propagators and vertices for QED is the same as for scalars. It is straightforward to generalize the results below by including the appropriate dirac and lorentz structure. This will be discussed in the next section.

We define the $n$-point functions in the 1-2 basis:
\bea
G^{(n)}(x_1,\cdots x_n)_{b_1 \cdots b_n}:= (-i)^{n-1} \langle {\cal P} [\phi(x_1)_{b_1} \cdots \phi(x_n)_{b_n}]\rangle  
\eea
The symbol ${\cal P}$ represents ordering along the closed time path. 
In what follows we will suppress the superscript ${(n)}$ and the co-ordinate variables and write the $n$-point function as $G_{b_1\cdots b_n}$.

\begin{figure}[H]
\begin{center}
\includegraphics[width=4cm]{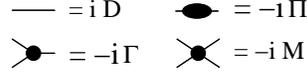}
\end{center}
\caption{Definitions of notation for propagator and vertices}
 \label{idefn}
\end{figure}

Vertex functions are obtained from the $n$-point functions by truncating external legs. 
In the 1-2 basis we write:
\begin{equation}
\label{trun}
G_{b_1\cdots b_n}= G_{b_1 \bar b_1} \cdots G_{b_n \bar b_n} 
	\Gamma^{\bar b_1\cdots \bar b_n}  \,.
\end{equation}
The vertex functions are obtained from the corresponding diagrams with an additional factor of $i$. 
This notation is illustrated schematically in Fig. \ref{idefn}.
\par
\xx In the 1-2 basis these functions satisfy the constraints:
\bea
&&\sum_{b_1=1}^2 \sum_{b_2=1}^2 \cdots \sum_{b_n=1}^2(-1)^{b_1+b_2+\cdots b_n + n}\;G_{b_1 b_2\cdots b_n}=0\\
&&\sum_{b_1=1}^2 \sum_{b_2=1}^2 \cdots \sum_{b_n=1}^2\Gamma_{b_1 b_2\cdots b_n}=0\nonumber
\eea

The rotation from the 1-2 representation to the Keldysh representation 
is accomplished by using the transformation matrix:
\begin{equation}
\label{firstU}
U_{Keldysh\leftarrow (1-2) }=\frac{1}{\sqrt{2}}\left(
\begin{array}{lr}
1 & 1 \\
1 & -1
\end{array}
\right).
\end{equation}
The $n$-point
function and vertex in the Keldysh representation are given by:
\bea
\label{Keldyn}
&& G_{c_1\cdots c_n}=2^{\frac{n}{2}-1}\, U_{c_1}\!^{b_1} 
\cdots U_{c_n}\!^{b_n} G_{b_1\cdots b_n}\\
&& \Gamma^{c_1\cdots c_n}=2^{\frac{n}{2}-1}\, U^{c_1}\!_{b_1} 
\cdots U^{c_n}\!_{b_n} \Gamma^{b_1\cdots b_n}\,.\nonumber
\eea
For the 2-point green function and vertex function in momentum space we use the notation
\bea
\label{kms}
&&G_{ra}(P)=G_{ret}(P)\,;~~G_{ar}(P)=G_{adv}(P)\,;~~G_{rr}(P)=G_{sym}(P) = N(P)(G_{ret}(P)-G_{adv}(P))\\
&&\Pi_{ar}(P)=\Pi_{ret}(P)\,;~~\Pi_{ra}(P)=\Pi_{adv}(P)\,;~~\Pi_{aa}(P)=\Pi_{sym}(P) = N(P)(\Pi_{ret}(P)-\Pi_{adv}(P))\nonumber
\eea
where $N(P)=1-2 n(p_0)$ and $n(p_0) = 1/(e^{\beta p_0}-1)$. The expression relating the symmetric function to the retarded and advanced functions is the KMS condition. 

In order to simplify the notation for the vertices, we replace each combination of the indices $\{r,a\}$ by a single numerical index. In momentum space we write:
\bea
\Gamma^{c_1 c_2\cdots c_n}(p_1,p_2, \cdots p_n) = \Gamma(i,p_1,p_2, \cdots p_n)
\eea 
We assign 
the choices of the variables $c_1 c_2 \cdots c_n$ to the variable $i$ using the vector
\bea 
\label{ralist}
V_n = 
\Big(
\begin{array}{c}
r_n \\
a_n
\end{array}
\Big) \cdots \otimes
\Big(
\begin{array}{c}
r_2 \\
a_2
\end{array}
\Big)\otimes
\Big(
\begin{array}{c}
r_1 \\
a_1
\end{array}
\Big)
\eea
where the symbol $\otimes$ indicates the outer product. For each $n$,
the $i$th component of the vector corresponds to a list of variables
that is assigned the number $i$.  To simplify the notation we drop the
subscripts and write a list like $r_1 r_2 a_3$ as $rra$. For clarity,
the results are listed below. \\

\xx 3-point functions: $rrr\ra 1$, $arr\ra 2$, $rar\ra 3$, $aar\ra 4$, $rra\ra 5$, $ara\ra 6$, $raa\ra 7$, $aaa\ra 
8$\\

\xx 4-point functions: $rrrr \ra 1$, $arrr \ra 2$,  $ rarr \ra 3$,  $ aarr \ra 4$,  $ rrar \ra 5$,  $ arar \ra 6$,  $ raar \ra 7$,  $aaar \ra 8$,  $ rrra \ra 9$,  $ arra \ra 10$,  $ rara \ra 11$,  $ aara \ra 12$,  $ rraa \ra 13$,  $araa \ra 
14$,  $ raaa \ra 15$, $ aaaa \ra 16$\\

\subsection{Propagators and Vertices}
\label{PandV}
In this section we define our notation for QED propagators and vertices. 
Greek letters from the beginning of the alphabet are dirac indices and greek letters from the end of the alphabet are lorentz indices. We use latin letters for 1-2 closed time path indices. 
We will need the fermion propagator, the photon propagator, the (two fermion - photon) 3-point vertex,  the three photon 3-point vertex,  and three different kinds of 4-point functions which couple four fermions, two fermions and two photons, and four photons. 
Of course, the three photon vertex and all 4-point vertices are zero at the tree level. We will work in the high temperature limit where the masses can be set to zero. 

The concept of `pinch singularities' plays an important role in the calculation of transport coefficients. 
The basic idea is that there is an 
infinite number of terms that all contribute at the same order because of the low frequency limit in the kubo formula (\ref{cond}). This limit produces pairs of retarded and advanced propagators which carry the same momenta. When integrating a term of the form $\int dp_0 \;G^{ret}(P)G^{adv}(P)$, the integration contour is `pinched' between poles on each side of the real axis, and the integral contains a divergence known as  a `pinch singularity.' These divergences are regulated by using resummed propagators which account for the finite width of thermal excitations. This procedure introduces extra factors of the coupling in the denominators which change the power counting. As a consequence, there is an infinite set of graphs which contain products of pinching pairs that all need to be resummed.

\subsubsection{Propagators}

In coordinate space the propagators are written:
\bea
&& S_{\alpha\beta}^{ab}(x_1,x_2)\,;~~D_{\mu\nu}^{ab}(x_1,x_2) 
\eea
In momentum space we will use the same expressions (to simplify the notation we do not introduce tilde's) with the coordinate variables $\{x_1,x_2\cdots x_n\}$ replaced by the momentum variables $\{p_1,p_2\cdots p_n\}$. Momenta are always taken to be incoming. Note that there is an overall delta function in momentum space so that, for example, the fermion propagator can be written:
\bea
&& S_{\alpha\beta}^{ab}(p_2):=S_{\alpha\beta}^{ab}(-p_2,p_2) 
\eea
For the fermion propagator we use the notation:
\bea
&&S^{-1}(P) = \Psl-\Sigma(P)\,;~~\Sigma(P)=\gamma_0 \Sigma_0(P)+\hat p^i \gamma^i \Sigma_s(P) \\[2mm]
&&S(P) = \frac{\Pslfrac-\Sigma(P)}{P^2-\frac{1}{2}\hat\Sigma(P)} \,;~~~~\hat\Sigma(P) := {\rm Tr}(\Psl \Sigma(P))\nonumber
\eea
We set $\Sigma$ to zero expect where it is needed to regulate a pinch singularity. We write:
\bea
\label{S-pinch}
&&S_{ret}(P) =\Psl\;G_{ret}(P):= \frac{\Psl}{P^2+i\,\Sign(p_0)\epsilon-\frac{1}{2}\hat\Sigma_{ret}(P)}  \\
&&S_{adv}(P) =\Psl\;G_{adv}(P):= \frac{\Psl}{P^2-i\,\Sign(p_0)\epsilon-\frac{1}{2}\hat\Sigma_{adv}(P)}  \nonumber\\
&&G_{ret}(P)\;G_{adv}(P) = -\frac{\rho(P)}{{\rm Im}\hat\Sigma(P)}\nonumber
\eea
where we have defined
\bea
\label{rho}
\rho(P) &&:= i\,d(P) := i\,(G_{ret}(P)-G_{adv}(P))\\[2mm]
&&:=\Sign(p_0)\Delta(P)\nonumber
\eea
There are no pinch terms involving photon propagators (see section \ref{Conductivity}) and therefore we can write the photon propagator (in the feynman gauge) as:
\bea
\label{DD}
&& D^{ret}_{\mu\nu}(P) =-g_{\mu\nu} G_{ret}(P) := - \frac{g_{\mu\nu}}{P^2+i\,\Sign(p_0)\epsilon}\\
&& D^{adv}_{\mu\nu}(P) =-g_{\mu\nu} G_{adv}(P) := - \frac{g_{\mu\nu}}{P^2-i\,\Sign(p_0)\epsilon}\nonumber
\eea

Following the notation of \cite{peskin} we write cut fermion lines and cut photon lines  as follows:
\bea
\label{cut-lines}
&&S^d_{\alpha\beta}(P)=S^{ret}_{\alpha\beta}(P)-S^{adv}_{\alpha\beta}(P) = \Psl\,d(P) \,;~~~~\Psl=\theta(p_0) u(P)\bar u(P)-\theta(-p_0)v(P)\bar v(P)\\[2mm]
&&D^d_{\mu\nu}(P) = D^{ret}_{\mu\nu}(P)-D^{adv}_{\mu\nu}(P) = -g_{\mu\nu}\,d(P)\,;~~~~g_{\mu\nu}= (-1)\;\epsilon_\mu^\lambda(P)\, \epsilon^{*\,\lambda}_\nu(P)\nonumber
\eea
For future use we also define the principle parts:
\bea
\label{prin}
&&{\rm Prin}(P) := \frac{1}{2}\big(G^{ret}(P)+G^{adv}(P)\big)\\
&&S^{prin}_{\alpha\beta}(P)=\frac{1}{2}\big(S^{ret}_{\alpha\beta}(P)+S^{adv}_{\alpha\beta}(P)\big) = \Psl {\rm Prin}(P)\nonumber\\
&&D^{prin}_{\mu\nu}(P)=\frac{1}{2}\big(D^{ret}_{\mu\nu}(P)+D^{adv}_{\mu\nu}(P)\big)= -g_{\mu\nu} {\rm Prin}(P) \nonumber
\eea
Note that to simplify the notation we use the same expressions $G_{ret}(P)$, $G_{adv}(P)$, $d(P)$ and ${\rm Prin}(P)$ in (\ref{S-pinch}), (\ref{DD}), (\ref{cut-lines}) and (\ref{prin}) for fermion and photon propagators. In any equation it will be clear if a given function refers to a photon of fermion propagator,  depending on whether the momentum variable corresponds to a fermion or a photon line.

\subsubsection{Vertices}


In coordinate space the vertices are written:
\bea
&& \Lambda_{\alpha\mu\beta}^{acb}(x_1,x_2,x_3)\,;~~\Omega_{\mu\nu\tau}^{abc}(x_1,x_2,x_3)\\
&& (M^{SS})^{ab;cd}_{\alpha\beta;\gamma\delta}(x_1,x_2;x_3,x_4)\,;~~(M^{SD})^{ab;cd}_{\alpha\beta;\mu\nu}(x_1,x_2;x_3,x_4)\,;~~(M^{DD})^{ab;cd}_{\mu\nu;\lambda\tau}(x_1,x_2;x_3,x_4)\nonumber
\eea
The two terms in the first line indicate the (two fermion - photon) 3-point vertex and the three photon 3-point vertex, respectively. The three terms in the second line are the four fermion 4-point vertex, the (two fermion - two photon) 4-point vertex, and the four photon 4-point vertex. The variables on each side of the semicolon indicate the legs that will join with a pinching pair of propagators. 
In section \ref{Conductivity} we show that the pinch terms involve two kinds of vertices: 3-point functions in which the momentum on one leg (which we take to be the middle leg) goes to zero, and 4-point functions which connect to two pairs of propagators with the same momenta. We write these vertices: 
\bea
\label{pinch-ver}
&&\Lambda_{\alpha\mu\beta}(j,P):= \lim_{Q\rightarrow 0}\Lambda_{\alpha\mu\beta}(j,-P-Q,Q,P) = \Lambda_{\alpha\mu\beta}(j,-P,0,P)\\
&&(M^{SS})_{\alpha\beta;\gamma\delta}(j,P,K):= \lim_{Q\rightarrow 0}(M^{SS})_{\alpha\beta;\gamma\delta}(j,-P,P+Q,-K-Q,K) = (M^{SS})_{\alpha\beta;\gamma\delta}(j,-P,P,-K,K)\nonumber
\eea
with similar expressions for the vertices $\Omega$, $M^{SD}$ and $M^{DD}$. 
In these expressions the index $j$ refers to the keldysh component of the vertex, as defined in section \ref{CTP}.

In the pinch limit, the spatial part (using $\mu = (0,z)$) of the (2 fermion - photon) vertex can be decomposed as:
\bea
\Lambda^z(j,P) = \big(A(j)\gamma_0 p_0 +B(j)\gamma^i\,p^i\big)\;\hat p^z+E(j)\gamma^z
\eea
It is easy to show that: 
\bea
\label{vert-decomp}
\Ksl\Lambda^z(j,K)\Ksl = \frac{1}{2}\Ksl\hat\Lambda^z(j,K)-K^2 \Lambda^z(j,K)\,;~~~~\hat\Lambda^z(j,K) := {\rm Tr}(\Ksl \Lambda^z(j,K))
\eea
For the conductivity we will need only the quantity $\hat\Lambda^z(j,K)$. 
In addition, we will show that we only need the trace of a pinched 4-point vertex:
\bea
\label{Mhat}
\hat M(P,K):={\rm Tr}\big(\Psl[- i {\bf M}(P,K)]\Ksl\big)
\eea
where the quantity ${\bf M}(P,K)$ is a particular combination of keldysh components and is defined in section \ref{keldBS}.

\subsubsection{Compactified Notation}
It is important to define a simplified notation, in order to avoid a proliferation of indices which would make equations almost unreadable. 

In section \ref{2PI} we will use a single numerical subscript to represent all continuous and discrete indices. For example: a photon field is written $A^a_{\mu}(x):=A_1$; 
the fermion propagator is written $S_{\alpha\beta}^{ab}(x_1,x_2):=S_{12}$, etc.
We also use an einstein convention in which a repeated index implies a sum over discrete variables and an integration over space-time variables.

In section \ref{Conductivity} we give expressions that result from summing over keldysh indices. 
Keldysh components and momentum variables are written explicitly. Traces are over dirac indices only.


\section{The 2PI formalism}
\label{2PI}

\subsection{Basic Formalism}
\label{basic2PI}

The partition function is defined as:
\begin{eqnarray}
\label{Z2PI}
&& \nonumber Z[J,\eta,\bar{\eta},C,B]=\int D[{\cal A} \Psi \bar{\Psi}] ~{\rm Exp}~\Big[
i   \bigg(
       S_{cl} +J_1 {\cal A}_1+ \bar{\eta}_1\Psi_1+\bar{\Psi}_1 \eta_1+\frac{1}{2}C_{12}{\cal A}_1 {\cal A}_2+
B_{12}{\Psi}_1\bar \Psi_2
        \bigg)\Big].
\end{eqnarray}
Recall that using our notation, the repeated index represents a summation over all discrete indices and also a integral over space and time variables. In the expression above, the time integral is carried out along the closed time path contour so that functional derivatives produce path ordered green functions.

The generating functional for connected fields is:
\begin{eqnarray}
\label{W2PI}
&&W[J,\eta,\bar{\eta},C,B]=-i \ {\rm Ln} Z[J,\eta,\bar{\eta},C,B]
\eea
Taking functional derivatives we obtain:
\bea
\label{W-relations}
&&\frac{\delta W}{\delta J_1}=A_1;~~\frac{\delta W}{\delta \bar{\eta}_1}=\psi_1; ~~\frac{\delta W}{\delta \eta_1}=- \bar{\psi}_1; \\
&&\frac{\delta W}{\delta C_{12}}=\frac{1}{2}\left(A_1 A_2+i D_{12}\right);~~
             \frac{\delta W}{\delta B_{12}}=\psi_1\bar \psi_2+i S_{12}. \nonumber       
\end{eqnarray}
where the connected propagators are defined as:
\bea
\label{prop-defn}
&&iD_{12} = \langle {\cal A}_1 {\cal A}_2\rangle - \langle {\cal A}_1\rangle \langle {\cal A}_2\rangle\,;~~\langle {\cal A}\rangle = A \\
&& iS_{12} = \langle \Psi_1 \bar\Psi_2\rangle - \langle \Psi_1\rangle \langle \bar\Psi_2\rangle
\,;~~\langle \bar\Psi\rangle =  \psi\nonumber
\eea

The QED 2PI effective action is obtained by taking the double legendre transform of $W[J,\eta,\bar{\eta},C,B]$ with respect to
the sources. The expression is explicitly constructed so that partial derivatives with respect to the sources are zero. We have:
\begin{eqnarray}
\label{Legendre}
&& \Gamma[\psi, \bar\psi, A, S, D]=W[J,\eta,\bar{\eta},C,B]-J_1A_1-\bar{\eta}_1\psi_1-\bar{\psi}_1 \eta_1-B_{12} (\psi_1 \bar{\psi}_2+i S_{12})- \frac{1}{2} C_{12}(A_1 A_2+ i D_{12})
\end{eqnarray}
By construction the effective action satisfies:
\begin{eqnarray}
\label{Gamma-relations}
&&\frac{\delta \Gamma}{\delta A_{1}}=-J_1-C_{12}A_2;~~
\frac{\delta \Gamma}{\delta \psi_1}=\bar{\eta}_1-B_{12} \bar{\psi}_2;~~
\frac{\delta \Gamma}{\delta \bar{\psi}_1}=-\eta_1+B_{21}\psi_2;\\
&&\frac{\delta \Gamma}{\delta D_{12}}=-\frac{i}{2}C_{12};~~
             \frac{\delta \Gamma}{\delta S_{12}}=-i B_{12}.\nonumber
\end{eqnarray}

Eqn (\ref{Legendre}) can be rewritten:
\begin{eqnarray}
\label{Gamma2PI}
&&\Gamma[\psi, \bar\psi, A, S, D]\\
&&=S_{cl}[\psi, \bar\psi, A]+
    \frac{i}{2} {\rm Tr} \,{\rm Ln}D^{-1}_{12}+
\frac{i}{2} {\rm Tr}\left[(D^0_{12})^{-1}\left(D_{21}-D^0_{21}\right)\right]-
  i {\rm Tr} \,{\rm Ln} S^{-1}_{12} - 
i{\rm Tr} \left[(S^0_{12})^{-1}(S_{21}-S^0_{21})\right]+\Phi[S,D] \nonumber
\eea
where $S_{cl}[\psi, \bar\psi, A]$ is the classical action and $S_0$ and $D_0$ are the free propagators given by:
\bea
\label{S0D0}
(S^0_{12})^{-1}=\frac{\delta^2S_{cl}}{\delta\psi_2\delta\bar\psi_1}\,;~~(D^0_{12})^{-1}=\frac{\delta^2 S_{cl}}{\delta A_2\delta A_1}\,,
\eea
The function $\Phi[S,D]$ is the sum of all 2PI diagrams. Note that for QED the function $\Phi[S,D]$ is independent of $\{\psi, \bar\psi, A\}$. The equations of motion of the mean field and the propagator are obtained from the stationarity of the action:
\bea
&&\frac{\delta \Gamma[\psi, \bar{\psi},A,S,D]}{\delta A}=0\,;~~\frac{\delta \Gamma[\psi, \bar{\psi},A,S,D]}{\delta D}=0\\
&&\frac{\delta \Gamma[\psi, \bar{\psi},A,S,D]}{\delta \psi}=0\,;~~\frac{\delta \Gamma[\psi, \bar{\psi},A,S,D]}{\delta \bar\psi}=0\,;~~\frac{\delta \Gamma[\psi, \bar{\psi},A,S,D]}{\delta S}=0\nonumber
\eea

\subsection{The Modified Effective Action}
\label{tildeGamma}

Practical calculations involve truncations, and it is well known that a straightforward truncation of $\Phi[S,D]$  leads to problems with gauge invariance \cite{AS,HZ}. 
The issue of gauge invariance can be addressed by introducing a different effective action defined with respect to the self-consistent solution of the propagator. We define $\tilde{S}[\psi,\bar{\psi},A]$ and $\tilde{D}[\psi,\bar{\psi},A]$ by:
\begin{eqnarray}
\label{constraint}
&& \frac{\delta \Gamma[\psi, \bar{\psi},A,S,D]}{\delta S}
  \bigg|_{S=\tilde{S}[\psi,\bar{\psi},A]} = 0\,; ~~~~ 
  \frac{\delta \Gamma[\psi, \bar{\psi},A,S,D]}{\delta D}
  \bigg|_{D=\tilde{D}[\psi,\bar{\psi},A]} = 0
\end{eqnarray}
These equations have the form of dyson equations:
\bea
\label{dyson}
&& \tilde S_{12}^{-1} = (S^0_{12})^{-1} - \Sigma_{12}\,;~~~~\Sigma_{12} = -i \frac{\delta \Phi[S, D]}{\delta S_{21}}\Big|_{\tilde S~\tilde D}\\
&& \tilde D_{12}^{-1} = (D^0_{12})^{-1} - \Pi_{12}\,;~~~~\Pi_{12} = 2 i \frac{\delta \Phi[S,D]}{\delta D_{21}}\Big|_{\tilde S~\tilde D}\nonumber
\eea
Substituting the self consistent solutions we 
obtain the modified action:
\begin{eqnarray}
\label{Gamma2PI-mod}
&& \tilde{\Gamma}[\psi,\bar{\psi},A]=
  \Gamma[\psi,\bar{\psi},A,\tilde{S}[\psi,\bar{\psi},A],\tilde{D}[\psi,\bar{\psi},A]]
\end{eqnarray}
The equivalence of (\ref{Gamma2PI}) and (\ref{Gamma2PI-mod}) at the exact level was shown in \cite{CJT}.

The external propagators are defined as 
\bea
\label{ext-prop}
(D^{{\rm ext}}_{12})^{-1}=   
     \frac{\delta^2}{\delta A_2 \delta A_1}
     \tilde{\Gamma}[\psi,\bar{\psi},A]\,;~~~~(S^{{\rm ext}}_{12})^{-1}=   
     \frac{\delta^2}{\delta \psi_2 \delta \bar\psi_1}
     \tilde{\Gamma}[\psi,\bar{\psi},A]
\eea
We also define the following vertex functions:
\begin{eqnarray}
\label{vert-defns}
&& \Lambda^0_{132} = -\frac{\delta (S^{0}_{12})^{-1}}{\delta A_3}\,;~~\Lambda_{132} = -\frac{\delta \tilde{S}^{-1}_{12}}{\delta A_{3}}\,;~~\Omega_{132} = -\frac{1}{2}\frac{\delta \tilde{D}^{-1}_{12}}{\delta A_{3}}\\
&&M^{S S}_{54;21}=-\frac{\delta^2 \Phi[\tilde S,\tilde D]}{\delta \tilde S_{12} \delta \tilde S_{45}}\,;~~
M^{S D}_{54;21} = -2\frac{\delta^2 \Phi[\tilde S,\tilde D]}{\delta \tilde S_{12} \delta \tilde D_{45}}\,;~~M^{D S}_{54;21} = -2\frac{\delta^2 \Phi[\tilde S,\tilde D]}{\delta \tilde D_{12}\delta \tilde S_{45} }\,;~~
M^{D D}_{54;21} = 4\frac{\delta^2 \Phi[\tilde S,\tilde D]}{\delta \tilde D_{12} \delta \tilde D_{45}}\nonumber
\end{eqnarray}
Some useful relations can be obtained from the identities:
\begin{eqnarray}
\label{in0}
\tilde{S}^{-1}_{13}\tilde{S}_{32}=\delta_{12}\,;~~\tilde{D}^{-1}_{13}\tilde{D}_{32}=\delta_{12}
\end{eqnarray}
Differentiating (\ref{in0}) with respect to $A$ and using (\ref{vert-defns}) gives:
\bea
\label{invert}
&& \frac{\delta\tilde{S}_{12}}{\delta A_{3}}= \tilde S_{11'}\Lambda_{1'3 2'}\tilde{S}_{2'2}\,;~~
\frac{\delta\tilde{D}_{12}}{\delta A_{3}} = 2
\tilde{D}_{11'}  \Omega_{1'32'}\tilde{D}_{2'2}
\eea

Taking derivatives of the dyson equations (\ref{dyson}) we obtain a set of bethe-salpeter type equations for the vertices. Using (\ref{vert-defns}) and (\ref{invert}) we get:
\begin{eqnarray}
\label{beta-sal}
\Lambda_{132}&&=
  -\frac{\delta}{\delta A_{3}}\left((S^{0}_{12})^{-1}-\Sigma_{12}\right)\\
&& =\Lambda^{0}_{132}-i\Big(
\frac{\delta^{2}\Phi[\tilde S,\tilde D]}{\delta \tilde S_{1'2'} \delta \tilde S_{21}}
    \frac{\delta \tilde{S}_{1'2'}}{\delta A_{3}}+ 
    \frac{\delta^{2}\Phi[\tilde S,\tilde D]}{\delta \tilde D_{1'2'} \delta \tilde S_{21}}
    \frac{\delta \tilde{D}_{1'2'}}{\delta A_{3}}\Big)\nonumber\\
&&= \Lambda^{0}_{132}+i\bigg(M^{S S}_{12;2'1'} [\tilde{S}_{1'4}\Lambda_{43 5} \tilde{S}_{52'}] 
+M^{S D}_{12;2'1'} [\tilde{D}_{1'4}\Omega_{43 5} \tilde{D}_{52'}]
  \bigg)\nonumber\\
\Omega_{132}&&=-\frac{1}{2}
  \frac{\delta}{\delta A_{3}}((D^{0}_{12})^{-1}-\Pi_{12})\nonumber\\
&&= i
  \left(
    \frac{\delta^{2}\Phi[\tilde S,\tilde D]}{\delta \tilde S_{1'2'} \delta \tilde D_{21}}
    \frac{\delta \tilde{S}_{1'2'}}{\delta A_{3}}+ 
    \frac{\delta^{2}\Phi[\tilde S,\tilde D]}{\delta \tilde D_{1'2'} \delta \tilde D_{21}}
    \frac{\delta \tilde{D}_{1'2'}}{\delta A_{3}}
  \right)\nonumber  \\ 
  &&= 
    -\frac{i}{2}\Big(
    M^{DS}_{12;2'1'} [\tilde{S}_{1'4}\Lambda_{43 5} \tilde{S}_{52'}]
    -M^{DD}_{12;2'1'} [\tilde{D}_{1'4}\Omega_{43 5} \tilde{D}_{52'}]
\Big)\nonumber
\end{eqnarray}
These equations are represented graphically in Fig. \ref{beta-sal-fig}.
\par\begin{figure}[H]
\begin{center}
\includegraphics[width=10cm]{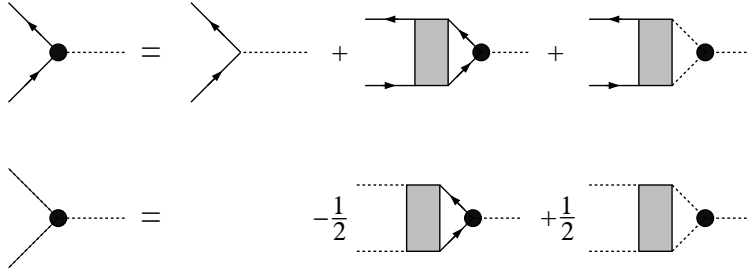}
\end{center}
\caption{Graphical representation of Eqn. (\ref{beta-sal})}
\label{beta-sal-fig}
\end{figure}

\subsection{The External Propagator}
\label{wi}

The external propagator can now be written in terms of these vertices. 
We will show below that these propagators satisfy the usual ward identities. The basic mechanism is simple:
the dyson equations (\ref{dyson}) contain $s$-channel ladder resummations and the bethe-salpeter equations (\ref{beta-sal}) introduce $t$- and $u$-channels, and thus restore the crossing symmetry.

We take the derivative of the modified effective action using the chain rule. In the expression below we suppress the arguments and write $\Gamma[\psi,\bar\psi,A, S, D]$ as $\Gamma$. We obtain:
\begin{eqnarray}
\label{Dext-long}
(D^{{\rm ext}}_{12})^{-1} &&
= \frac{\delta^2\Gamma}{\delta A_{2} \delta A_{1}}
+  \frac{\delta^2 \Gamma}{\delta D_{34}\delta D_{56}}\Big|_{\tilde S\;\tilde D}
   \frac{\delta \tilde{D}_{34}}{\delta A_{2}}
   \frac{\delta \tilde{D}_{56}}{\delta A_{1}}
+
   \frac{\delta^2 \Gamma}{\delta S_{34} \delta S_{56}}\Big|_{\tilde S\;\tilde D}
   \frac{\delta \tilde{S}_{34}}{\delta A_{2}}
   \frac{\delta \tilde{S}_{56}}{\delta A_{1}}\\
&&
+ \left(\frac{\delta^2 \Gamma}{\delta D_{34}\delta A_{1}}\Big|_{\tilde S\;\tilde D}
     \frac{\delta \tilde{D}_{34}}{\delta A_{2}}
+   \frac{\delta^2 \Gamma}{\delta S_{34}\delta A_{1}}\Big|_{\tilde S\;\tilde D}
   \frac{\delta \tilde{S}_{34}}{\delta A_{2}}
+
\frac{\delta^2 \Gamma}{\delta S_{34} \delta D_{56}}\Big|_{\tilde S\;\tilde D}
   \frac{\delta \tilde{D}_{34}}{\delta A_{2}}
   \frac{\delta \tilde{S}_{56}}{\delta A_{1}} ~+~\{1\leftrightarrow 2\}\right)
\nonumber
\end{eqnarray}  
Using (\ref{Gamma2PI}), (\ref{S0D0}), (\ref{vert-defns}) and (\ref{invert}) this expression can be rewritten. We give the result separately for each term:
\bea
\label{Dext-long2}
\frac{\delta^2\Gamma}{\delta A_{2} \delta A_{1}}&& =(D^{0}_{12})^{-1}\\
\frac{\delta^2 \Gamma}{\delta D_{34}\delta D_{56}}\Big|_{\tilde S\;\tilde D}
   \frac{\delta \tilde{D}_{34}}{\delta A_{2}}
   \frac{\delta \tilde{D}_{56}}{\delta A_{1}}
&& = 2i\Omega_{413} [\tilde D_{33'} \Omega_{3'24'} \tilde D_{4'4}]
+ M^{DD}_{65;43} [\tilde D_{55'}\Omega_{5'16'}\tilde D_{6'6}] [\tilde D_{33'}\Omega_{3'24'}\tilde D_{4'4}]\nonumber\\
\frac{\delta^2 \Gamma}{\delta S_{34} \delta S_{56}}\Big|_{\tilde S\;\tilde D}
   \frac{\delta \tilde{S}_{34}}{\delta A_{2}}
   \frac{\delta \tilde{S}_{56}}{\delta A_{1}}
&& = -i\Lambda_{413} [\tilde S_{33'} \Lambda_{3'24'} \tilde S_{4'4}]
- M^{SS}_{65;43} [\tilde S_{55'}\Lambda_{5'16'}\tilde S_{6'6}] [\tilde S_{33'}\Lambda_{3'24'}\tilde S_{4'4}]\nonumber\\
\frac{\delta^2 \Gamma}{\delta D_{34}\delta A_{1}}\Big|_{\tilde S\;\tilde D}
     \frac{\delta \tilde{D}_{34}}{\delta A_{2}} ~+~\{1\leftrightarrow 2\} && = 0\nonumber\\
\frac{\delta^2 \Gamma}{\delta S_{34}\delta A_{1}}\Big|_{\tilde S\;\tilde D}
   \frac{\delta \tilde{S}_{34}}{\delta A_{2}^{\nu}} ~+~\{1\leftrightarrow 2\}
&&  = i\Lambda^0_{413} [\tilde S_{33'} \Lambda_{3'24'} \tilde S_{4'4}]~+~\{1\leftrightarrow 2\}\nonumber\\
\frac{\delta^2 \Gamma}{\delta S_{34} \delta D_{56}}\Big|_{\tilde S\;\tilde D}
   \frac{\delta \tilde{D}_{34}}{\delta A_{2}}
   \frac{\delta \tilde{S}_{56}}{\delta A_{1}} ~+~\{1\leftrightarrow 2\}
&& = - M^{SS}_{65;43} [\tilde S_{55'}\Lambda_{5'16'}\tilde S_{6'6}] [\tilde D_{33'}\Omega_{3'24'}\tilde D_{4'4}]~+~\{1\leftrightarrow 2\}\nonumber
\eea
The complicated set of terms above can be expressed in a simple form by using the bethe-salpeter equations. Substituting  (\ref{beta-sal}) into (\ref{Dext-long2}) we arrive at the following compact result for the external propagator:
\begin{eqnarray}
\label{dext}
&&  (D^{\rm ext}_{12})^{-1}=(D^{0}_{12})^{-1}+
   i(\Lambda^{0}_{314}\tilde{S}_{44'}\Lambda_{4'23'}\tilde{S}_{3'3})
\end{eqnarray}
From (\ref{dext}) we extract the vertex part of the 2-point function:
\bea
\label{piFirst}
\Pi^{{\rm ext}}_{12}=-i(\Lambda^{0}_{314}\tilde{S}_{44'}\Lambda_{4'23'}\tilde{S}_{3'3}):=-i {\rm Tr}\,[\Lambda^{0}_1\tilde{S}\Lambda_2\tilde{S}]
\eea
This result is illustrated in Fig. \ref{piF}. The trace is over the indices that correspond to the closed loop.
\par\begin{figure}[H]
\begin{center}
\includegraphics[width=5cm]{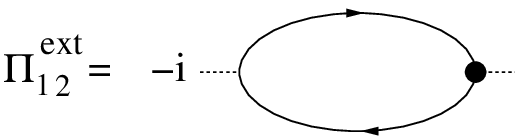}
\end{center}
\caption{Graphical representation of Eqn (\ref{piFirst})}
 \label{piF}
\end{figure}

It is straightforward to show that the external propagator (\ref{ext-prop}) satisfies the usual ward identity:
\bea
\partial_1 (D_{12}^{\rm ext})^{-1} = 0
\eea
We use the fact that the action and the integral measure are invariant under the transformation:
\begin{eqnarray}
\label{gauge-trans}
&& {\cal A}_1\rightarrow {\cal A}_1+\partial_1\Lambda_1\,;~~ 
    \Psi_1\rightarrow \Psi_1-i e    \Lambda_1\Psi_1\,;~~\bar{\Psi}_1\rightarrow \bar{\Psi}_1+i e \Lambda_1\bar{\Psi}_1~~~{\rm no~sum}
\end{eqnarray}
We work to first order in $\Lambda$ and integrate by parts where necessary so that we can extract an overall factor of $\Lambda$. We obtain:
\begin{eqnarray}
\label{wi1}
&&  \Delta Z[J,\eta, \bar{\eta},C,B] \\
&& = \int D[{\cal A} \Psi \bar{\Psi}]~
\bigg[
-\Lambda_1 \partial_{1}J_1 - i e \Lambda_1(\bar{\eta}_1\Psi_1-\bar\Psi_1\eta_1 ) -\Lambda_1\partial_1 C_{12}{\cal A}_2
-ieB_{12}(\Lambda_1-\Lambda_2)\Psi_1\bar\Psi_2\bigg] \nonumber\\
&&~~~~~~~~\cdot~{\rm Exp}~\Big[
i   \bigg(
       \int dx\;L +J_1 {\cal A}_1+ \bar{\eta}_1\Psi_1+\bar{\Psi}_1 \eta_1+\frac{1}{2}C_{12}{\cal A}_1 {\cal A}_2+
B_{12}{\Psi}_1\bar \Psi_2
        \bigg)\Big] = 0\nonumber
\eea
We rewrite this expression using the standard trick to extract the quantum fields from the path integral. We replace the fields by derivatives with respect to the sources, which act on the exponential factor in (\ref{wi1}):
\bea
{\cal A}_1\rightarrow \frac{1}{i}\frac{\delta}{\delta J_1}\,;~~\Psi_1\rightarrow \frac{1}{i}\frac{\delta}{\delta \bar\eta_1}\,;~~\bar\Psi_1\rightarrow -\frac{1}{i}\frac{\delta}{\delta \eta_1}\,;~~\Psi_1 \bar\Psi_2 \rightarrow \frac{1}{i}\frac{\delta}{\delta B_{12}}\,;~
\eea
Using (\ref{W2PI}) and (\ref{W-relations}) we can rewrite the derivatives with respect to the sources in terms of the expectation values of the fields. 
We obtain:
\bea
-\Lambda_1\partial_1 J_1-ie\Lambda_1(\bar \eta_1\psi_1-\bar\psi_1\eta_1)-\Lambda_1\partial_1 C_{12}A_2-ie B_{12}(\Lambda_1-\Lambda_2)(\psi_1\bar\psi_2+i S_{12}) = 0
\eea
We rewrite the above expression using (\ref{Gamma-relations}) to replace the sources by the appropriate derivatives of the effective action. The result is:
\bea
\label{differ1}
\Lambda_1 \partial_1\frac{\delta \Gamma}{\delta A_1}-ie\Lambda_1
\left(\bar\psi_1\frac{\Gamma}{\delta \bar\psi_1}+\frac{\Gamma}{\delta \psi_1}\psi_1\right)
+i e \frac{\delta \Gamma}{\delta S_{12}}(\Lambda_1-\Lambda_2)S_{12} =0
\eea
This result can be rewritten as a total differential by using (\ref{gauge-trans}) to obtain
\bea
\label{gaugeDEL}
\Delta A_1=\partial_1\Lambda_1 \,;~~\Delta\psi_1=-i e    \Lambda_1\psi_1 \,;~~ \Delta\bar\psi_1=i e    \Lambda_1\bar\psi_1 \,;~~i \Delta S_{12} = -i e (\Lambda_1-\Lambda_2)i S_{12}\,;~~i \Delta D_{12}=0 ~~~~{\rm no~sum}
\eea
where the last two equations come directly from (\ref{prop-defn}). From (\ref{differ1}) we get:
\bea
\Delta\Gamma = \Delta A_1\frac{\delta \Gamma}{\delta A_1}+\Delta D_{12}\frac{\delta \Gamma}{\delta D_{12}}+\Delta \bar\psi_1\frac{\delta \Gamma}{\delta \bar\psi_1}+\Delta \psi_1\frac{\delta \Gamma}{\delta \psi_1}+\Delta S_{12}\frac{\delta \Gamma}{\delta S_{12}}=0
\eea
and thus, independent of the truncation scheme, we have: 
\bea
&&\Delta\tilde\Gamma = \Delta A_1\frac{\delta \tilde\Gamma}{\delta A_1}+\Delta \bar\psi_1\frac{\delta \tilde\Gamma}{\delta \bar\psi_1}+\Delta \psi_1\frac{\delta \tilde\Gamma}{\delta \psi_1}=0
\eea
Using (\ref{gaugeDEL}) we obtain
\bea
\label{wi-general}
\partial_1\frac{\delta\tilde\Gamma}{\delta A_1}-ie\left(\bar\psi_1\frac{\delta\tilde\Gamma}{\delta\bar\psi_1}+\frac{\delta\tilde\Gamma}{\delta\psi_1}\psi_1\right)=0
\eea
We get ward identities by taking functional derivatives of (\ref{wi-general}) with respect to the fields, and setting the remaining fields to zero. For example, differentiating with respect to $A_2$ and setting $\{A,\psi,\bar\psi\}$ to zero gives:
\bea
\partial_1\frac{\delta^2 \tilde\Gamma}{\delta A_1\delta A_2} = \partial_1(D^{\rm ext}_{12})^{-1}=0
\eea

\subsection{Bethe-Salpeter Equation}
\label{BetaSalEqn}

We obtain a bethe-salpeter type integral equation for the (two fermion - photon) 3-point vertex by substituting the second equation in Eqn (\ref{beta-sal}) into the first and iterating. 
We show the result in Fig. \ref{bSx} below.
\par\begin{figure}[H]
\begin{center}
\includegraphics[width=8cm]{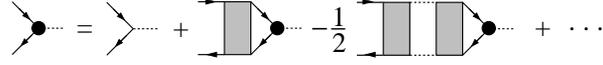}
\end{center}
\caption{Graphical representation of Eqn (\ref{bs4})}
 \label{bSx}
\end{figure}
The corresponding equation can be written:
\bea
\label{bs4}
\Lambda_{132} &&=\Lambda^{0}_{132}+i {\cal M}^{SS}_{12;45} \tilde S_{55'}\Lambda_{5'34'}\tilde S_{4'4}
\eea
where we have defined the composite vertex:
\bea
\label{M1}
&&{\cal M}^{SS}_{12;45}=M^{SS}_{12;45}-\frac{i}{2}M^{SD}_{12;2'1'}\tilde D_{1'6}\tilde D_{7,2'}M^{DS}_{67;45}
\eea
To reduce the number of indices, we introduce a type of matrix notation. The indices that correspond to legs that are joined in a closed loop are summed over, and are not written explicitly. Using this notation Eqn (\ref{M1}) is written: 
\bea
\label{Mf}
{\cal M}^{SS}_{12;45}=M^{SS}_{12;45}-\frac{i}{2}\Big(M^{SD}~\tilde D~\tilde D ~M^{DS}\Big)_{12;45}
\eea
and Eqn (\ref{bs4}) becomes:
\bea
\label{bs5}
\Lambda_{132} &&=\Lambda^{0}_{132}+i \Big({\cal M}^{SS} \tilde S\;\tilde S\;\Lambda \Big)_{132} 
\eea
Equation (\ref{bs5}) is illustrated in Fig. \ref{bSy}. The dark box indicates the composite vertex ${\cal M}$ defined in (\ref{Mf}). For clarity, the indices that are not summed over are shown explicitly on the diagram. 
\par\begin{figure}[H]
\begin{center}
\includegraphics[width=8cm]{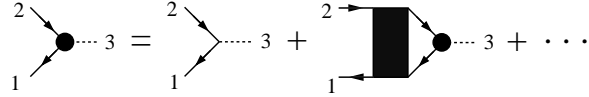}
\end{center}
\caption{Graphical representation of Eqn (\ref{bs5})}
\label{bSy}
\end{figure}
\noindent We note that this notation does not really represent matrix multiplication because of the fact that the indices which are summed over cannot always be written next to each other. However, the meaning of the notation is immediately clear from the corresponding diagram.

The vertex ${\cal M}$ is obtained from (\ref{vert-defns}) and (\ref{Mf}).  In order to include all contributions that correspond to binary scattering and production processes within the 2PI formalism, we need to work to 3-loop order in the $\Phi$ functional. The two diagrams we need are shown in Fig. \ref{Phi}. 
\par\begin{figure}[H]
\begin{center}
\includegraphics[width=4cm]{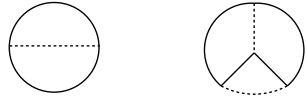}
\end{center}
\caption{2PI contributions to the $\Phi$ functional}
\label{Phi}
\end{figure}
\noindent The first graph in Fig. \ref{Phi} produces the leading order contributions to the vertices and the second graph gives next-to-leading order terms. There are no next-to-leading order contributions to the vertices $M^{SD}$ and $M^{DS}$. We write 
\bea
\label{Mexp}
M^{SS}=M^{SS}_{lo}+M^{SS}_{nlo}\,;~~~~M^{SD}=M^{SD}_{lo}\,;~~~~M^{DS}=M^{DS}_{lo}
\eea
The results are shown  in Fig. \ref{sk-vert}.
\par\begin{figure}[H]
\begin{center}
\includegraphics[width=9cm]{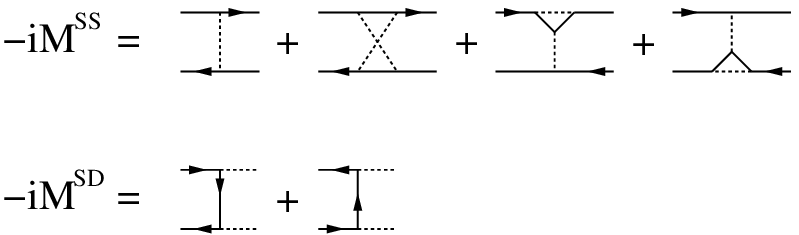}
\end{center}
\caption{Contributions to the 4-point functions}
\label{sk-vert}
\end{figure}
\noindent Substituting (\ref{Mf}) and (\ref{Mexp}) into (\ref{bs5}) we iterate and keep terms that correspond to binary scattering and production processes. We obtain: 
\bea
\label{bs7}
\Lambda_{132} &&=\Lambda^{0}_{132}+\Big(\underbrace{i M^{SS}_{lo}}_a +\underbrace{i M^{SS}_{nlo}}_{\{c,d,e\}} -\underbrace{M_{lo}^{SS}\,\tilde S\,\tilde S M_{lo}^{SS}}_{b} + \frac{1}{2}\underbrace{M_{lo}^{SD}\,\tilde D\,\tilde D \,M_{lo}^{DS}}_{\{f,g\}}\Big)_{12;45}\,\tilde S_{55'}  \,\tilde S_{4'4}\,\Lambda_{5'34'}
\eea
Equation (\ref{bs7}) is shown in Fig. \ref{bSz}.
\par\begin{figure}[H]
\begin{center}
\includegraphics[width=8cm]{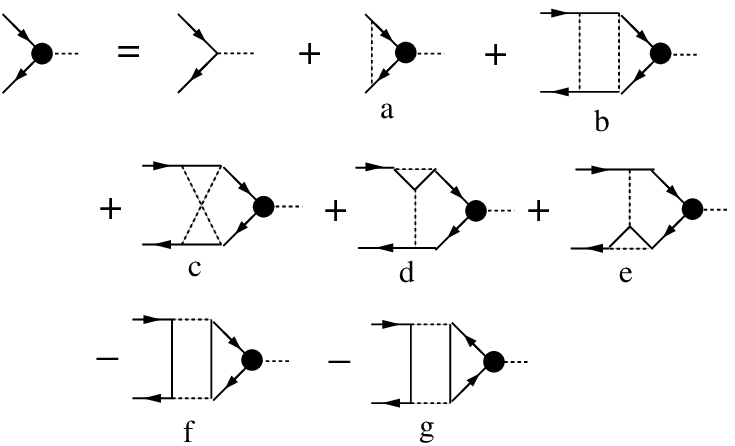}
\end{center}
\caption{Binary contributions to the bethe-salpeter equation}
 \label{bSz}
\end{figure}
We write the final result:
\bea
\label{bs8}
\Lambda_{132} &&=\Lambda^{0}_{132}~~+ \sum_{j\in \{a,b,c,d,e,f,g\}}i\,M^{(j)}_{12;45}\,\tilde S_{55'}  \,\tilde S_{4'4}\,\Lambda_{5'34'}
\eea
the superscript $j\in \{a,b,c,d,e,f,g\}$ indicates the 4-point function shown in part $(j)$ of Fig. \ref{bSz}.

\section{Electrical Conductivity}
\label{Conductivity}

\subsection{Kubo Formula}
\label{kubo}
We introduce several notational changes in this section. Keldysh components and momentum variables are written explicitly, and  traces are over dirac indices only. All tilde's on propagators are suppressed.  We also introduce a slightly different notation for the vertices. No factors of the coupling have been extracted from the vertices as defined in (\ref{vert-defns}). For example, using (\ref{S0D0}) and (\ref{vert-defns}) gives $\Lambda^0_\mu=e\,\gamma_\mu$. In the rest of this section we extract one power of the coupling from both $\Lambda^0_\mu$ and $\Lambda_\mu$ to obtain expressions with conventional form. To simplify the notation, we do not introduce additional primes on the new vertices.

The electrical conductivity can be obtained from the kubo formula:
\bea
\label{cond}
&&\sigma=\frac{1}{6}\left(\frac{\partial}{\partial q_0}2\, {\rm Im}\,\rho^{ii}(q_0,0)\right)\Big|_{q_0\rightarrow 0}\\[2mm]
&&\rho^{ii}(x,y)=\langle j^i(x)j^i(y)\rangle\,;~~j^i(x)=\bar\psi(x)\gamma^i\psi(x)\nonumber
\eea
We can write the conductivity in terms of the polarization tensor using:
\bea
\rho^{ii}(q_0,0)=-\frac{1}{e^2}\Pi^{ii}_{ret}(q_0,0)
\eea

The polarization tensor is given in (\ref{piFirst}). Summing over keldysh indices we obtain:
\bea
\label{condx}
\Pi^{ret}_{\mu\nu}(Q)&&=\frac{i}{2}e^2\int dP\;{\rm Tr}\big(\Lambda^0_\mu \big[S^{sym}(P) \Lambda_\nu(5,-P,-Q,P+Q) S^{ret}(P+Q)\\
&&+S^{adv}(P) \Lambda_\nu(2,-P,-Q,P+Q) S^{sym}(P+Q)+S^{adv}(P) \Lambda_\nu(6,-P,-Q,P+Q) S^{ret}(P+Q)\big]\big)\nonumber
\eea
Note that the surviving indices on the 3-point functions correspond to the middle legs of the 3-point vertices.  
Eqn. (\ref{condx}) can be rewritten using the kms conditions. The kms condition for the 2-point function is given in (\ref{kms}). Similar kms conditions exits for the 3-point functions. A complete list is given in \cite{MCTF}. The expression we need in this case is:
\bea
\label{kms3}
\Gamma(6,P_1,P_2,P_3)=N_F(p^0_3)\big(\Gamma^*(3,P_1,P_2,P_3)-\Gamma(2,P_1,P_2,P_3)\big)+N_F(p^0_1)\big(\Gamma^*(3,P_1,P_2,P_3)-\Gamma(5,P_1,P_2,P_3)\big)
\eea
Substituting (\ref{kms}) and (\ref{kms3}) into (\ref{condx}) and only including terms which contain pinch singularities gives:
\bea
\label{condy}
\Pi^{ret}_{\mu\nu}(Q)=-\frac{i}{2}e^2\int dP\;(N_F(p_0)-N_F(p_0+q_0)) {\rm Tr}\big(\Lambda^0_\mu S^{adv}(P) \Lambda^*_\nu(3,-P,-Q,P+Q) S^{ret}(P+Q)\big)
\eea
Substituting (\ref{condy}) into (\ref{cond}) we obtain:
\bea
\label{piNext}
\sigma = \frac{1}{3} \beta\int dP\;(1-n_f(p_0))n_f(p_0){\rm Tr}\big(\Lambda_0^i S^{ret}(P) {\rm Re}\Lambda^i(3,P) S^{adv}(P)\big)
\eea
As explained in section \ref{CTP} the index `3' indicates the keldysh component that is retarded with respect to the middle leg. 
We substitute (\ref{S-pinch}) and (\ref{vert-decomp}) into (\ref{piNext}) and obtain:
\bea
\label{piLast}
\sigma = \frac{4}{3}\beta\int dP\;(1-n_f(p_0))n_f(p_0)\rho(P) p^i B^i(P)\,;~~~~B^i(P):=\frac{{\rm Re}\hat\Lambda^i(3,P)}{2{\rm Im}\hat\Sigma(P)}
\eea
In the next three sections we obtain a self-consistent integral equation for the quantity $B^i(P)$ in (\ref{piLast}). We will show that this integral equation has the same form as that obtained in \cite{AMY} using kinetic theory. 

\subsection{Bethe-Salpeter Equation}
\label{keldBS}

From Eqn (\ref{piLast}) we only need to calculate the third keldysh component of the 3-point vertex in (\ref{bs8}).
We do the sums over keldysh indices and only include terms which contain pinch singularities. Eqn (\ref{bs8}) becomes:
\bea
\label{term1}
\Lambda_{\alpha\mu\beta}(3,P) =  \Lambda^0_{\alpha\mu\beta}(3,P)~~+ \sum_{j\in \{a,b,c,d,e,f,g\}}\frac{i}{2}\int dK\,{\bf M}_{\alpha\beta;\gamma\delta}^{(j)}(P,K) S_{\delta\delta'}^{ret}(K) \Lambda_{\delta'\mu\gamma'}(3,K)  S_{\gamma'\gamma}^{adv}(K) 
\eea
where the 4-point function that appears in this equation is a combination of keldysh components and thermal functions:
\bea
\label{Mfkeld}
&& {\bf M}^{(j)}(P,K) = M^{(j)}(13,P,K)+N_F(K)\Big(M^{(j)}(5,P,K)-M^{(j)}(9,P,K)\Big)\nonumber
\eea
The superscript $j\in \{a,b,c,d,e,f,g\}$ indicates the 4-point function shown in part $(j)$ of Fig. \ref{bSz}.
From (\ref{piLast}) we need to find an integral equation for $\hat\Lambda^i(P) = {\rm Tr}\big(\Psl \Lambda^i(P)\big)$. We multiply both sides of (\ref{term1}) by $\Psl$ and take the trace. Using (\ref{S-pinch}), (\ref{vert-decomp}) and (\ref{Mhat}) we obtain:
\bea
\label{bSw}
&&\hat\Lambda^i(3,P) = \hat\Lambda_0^i(3,P)~~+ \sum_{j\in \{a,b,c,d,e,f,g\}}\frac{1}{2}\int dK\;\hat M^{(j)}(P,K) \rho(K)B^i(K)
\eea
Taking the real part of both sides and using the definition of $B^i(P)$ in (\ref{piLast}) we obtain:
\bea
\label{AMY0}
2 {\rm Im}\,\hat \Sigma(P)\cdot B^i(P) = {\rm Re}\,\hat\Lambda_0^i(3,P)~~+ \sum_{j\in \{a,b,c,d,e,f,g\}}\frac{1}{2}\int dK\;{\rm Re}\,\big[\,\hat M^{(j)}(P,K) \,\big]\rho(K)B^i(K)
\eea
 
\subsection{Matrix Elements}
\label{ME}

\subsubsection{Preliminaries}

In this section we discuss the structure of each of the factors ${\rm Re}\,\big[\,\hat M^{(j)}(P,K) \,\big]$ in (\ref{AMY0}).
We show that the complete set of diagrams produces all of the amplitudes that correspond to binary scattering and production processes. The calculation for each diagram is similar. We outline the procedure below. In Appendix A we give the details  for two diagrams: $\hat M^{(b)}(P,K)$ and $\hat M^{(d)}(P,K)$.\\

\xx (1) The first step is to sum over the keldysh indices. These sumations can be done by hand, but the calculation is extremely tedious. Instead, we use a Mathematica program. This program is described in detail in  \cite{MCTF} and is available at www.brandonu.ca/physics/fugleberg/Research/Dick.html. The program can be used to calculate the integrand corresponding to any diagram (up to five external legs) in the keldysh, RA or 1-2 basis. The user supplies input in the form of lists of momenta and vertices for each propagator and vertex. \\

\xx (2) The second step is to divide the result into real and imaginary parts.  The method is related to the Cutkosky rules at finite temperature, and is described in \cite{MC2}. In Appendix A we explain in detail how the procedure works for two examples.\\

\xx (3) The real part of each diagram will contain two cut internal lines. Opening these two lines effectively divides each diagram into two pieces. The last step in the calculation is to combine the pieces from all diagrams and show that the total result can be written as the square of the amplitudes that correspond to binary processes. The procedure is as follows. For the moment, we call the momenta of the internal cut lines $R_1$ and $R_2$.
In principle, each diagram $\hat M^{(j)}(P,K)$ contains 16 terms which correspond to the $2^4$ possible choices for the signs of the 0-components of the momenta on the four on-shell lines: $\{p^0,\;k^0,\;r_1^0,\;r_2^0\}$. Since $P$ is an external variable in the integral equation (\ref{AMY0}), we make the choice $p^0>0$, which leaves eight terms. Only three of these terms correspond to kinematically allowed binary processes, or 2 $\to$ 2 scattering/production processes. 

To write the matrix elements in conventional form, we rewrite the four variables $\{P,\;K,\;R_1,\;R_2\}$ in terms of the new variables $\{P,\;P_2,\;L_1,\;L_2\}$ which are defined so that $P+P_2=L_1+L_2$ and $\Sign (p^0)=\Sign (p_2^0)=\Sign (l_1^0)=\Sign (l_2^0)$. Using this notation, the expression for $\hat M^{(j)}(P,K)$ will contain an overall factor $\int dL_1\int dL_2 \;\delta_{P+P_2-L_1-L_2}:=\int dL_1\int dL_2 \;\delta^4\,(P+P_2-L_1-L_2)$. 

For each diagram we have labeled the momenta so that the internal cut lines carry momenta $\{P-L,K\pm L\}$  (see Figs. \ref{diagB} to \ref{diagCv}). 
In each case, there are three ways to define the variables $\{P,\;P_2,\;L_1,\;L_2\}$   corresponding to the three possible ways to select $P_2$ from the set $\{\pm K,\;\pm (P-L),\;\pm(K\pm L)\}$.
For the case of diagrams (b) and (d), the cut lines carry momenta $\{P-L,K-L\}$ and the three possible choices are:
\bea
\label{scat-vars}
&& (1)~~~~P_2=-K\,;~~\{L_1,\;L_2\}=\{P-L,\;-(K-L)\} \\
&& (2)~~~~P_2=K-L\,;~~\{L_1,\;L_2\}=\{K,\;P-L\} \nonumber\\
&& (3)~~~~P_2=-(P-L)\,;~~\{L_1,\;L_2\}=\{K,\;-(K-L)\}\nonumber
\eea
For diagrams for which the cut lines carry momenta $\{P-L,K+L\}$ the three possible choices are:
\bea
\label{scat-vars2}
&& (1)~~~~P_2=K\,;~~\{L_1,\;L_2\}=\{P-L,\;K+L\} \\
&& (2)~~~~P_2=-(K+L)\,;~~\{L_1,\;L_2\}=\{P-L,\;-K\} \nonumber\\
&& (3)~~~~P_2=-(P-L)\,;~~\{L_1,\;L_2\}=\{K+L,\;-K\} \nonumber
\eea
In both cases, the terms corresponding to choice (2) and choice (3) can be obtained from the term corresponding to choice (1) by making the changes of variables: $P_2\leftrightarrow -L_1$ and $P_2\leftrightarrow -L_2$. 
We define the notation
\bea
\label{defn-perms}
&&\int dP_2\int dL_1\int dL_2\;\delta_{P+P_2-L_1-L_2}~\sum_{perms}f(P,P_2;L_1,L_2) \\
&&= \int dP_2\int dL_1\int dL_2\;\delta_{P+P_2-L_1-L_2}~\bigg(f(P,P_2;L_1,L_2)
+f(P,-L_1;-P_2,L_2)+f(P,-L_2;L_1,-P_2)\bigg)\nonumber
\eea
\\

\xx (4) Finally, we need to show that the thermal factors for each term are correct. Each term should be weighted with a factor that corresponds to the product of the appropriate statistical emission and absorption factors. This result is obtained by using the identity
\bea
\label{N-identity}
1+N_B(p^0_1) N_F(p^0_2)+N_F(p^0_3) N_F(p^0_2)+N_B(p^0_1) N_F(p^0_3) = 0~~{\rm if}~~ p^0_1+p^0_2+p^0_3= 0
\eea
where we have used the usual definitions:
\bea
&&N_B(p_0) = 1-2n_b(p_0)\,;~~~n_b(p_0) = \frac{1}{e^{\beta p_0}-1}\\
&&N_F(p_0) = 1+2n_f(p_0)\,;~~~n_f(p_0) = \frac{1}{e^{\beta p_0}+1}\nonumber
\eea
For diagrams (b) and (d) the details are given in Appendix A. The result is that the thermal factor for each diagram has the form of one of the two expressions below:
\bea
\label{Ncal}
&& {\cal N}_f = n_f(p^0)(1-n_f(l^0_1))(1-n_f(l^0_2)) \\
&& {\cal N}_b = n_f(p^0)(1-n_b(l^0_1))(1-n_b(l^0_2))\nonumber
\eea

\subsubsection{Diagram (b)}
\label{b}

We begin by looking at diagram (b) which is shown on the left hand side of Fig. \ref{diagB}. The real part of $\hat M^{(b)}(P,K)$  can be written as the product of the amplitudes shown in the right hand side of Fig. \ref{diagB}.  
\par\begin{figure}[H]
\begin{center}
\includegraphics[width=8cm]{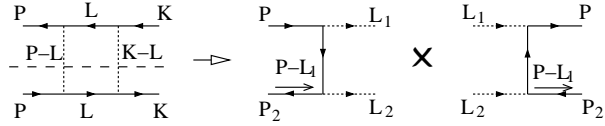}
\end{center}
\caption{The real part $\hat M^{(b)}$ produces $|m^t_{e^+ e^-\rightarrow \gamma\gamma}|^2$}
 \label{diagB}
\end{figure}
Using the first line in (\ref{scat-vars}) we obtain (see Appendix A):
\bea
&&(1-n_f(p^0))\cdot{\rm Re}\,\big[\,\hat{M}^{b}\,\big] \\
&&~~~~= 2 \sum_{perms}\int dL_1\int dL_2\;\delta_{P+P_2-L_1-L_2}\;{\cal N}_b \;({\bf m}_b^\dagger)^{ss'\rightarrow \lambda\lambda'}\cdot ({\bf n}_b)^{ss'\rightarrow \lambda\lambda'}~\Delta(L_1)\;\Delta(L_2)\nonumber\\
&&~~~~ ({\bf m}^\dagger_b)^{ss'\rightarrow \lambda\lambda'}=({\bf n}^\dagger_b)^{ss'\rightarrow \lambda\lambda'} = e^2\,\bar u_\alpha^s(P)\,\big(\gamma^\mu S^{ret}(P-L_1)\gamma^\nu\big)_{\alpha\delta}\;v_\delta^{s'}(P_2)\;\epsilon_\mu^\lambda(L_1)\;\epsilon^{\lambda'}_\nu(L_2)\nonumber
\eea
The amplitude $({\bf m}_b)^{ss'\rightarrow \lambda\lambda'}$ corresponds to the $t$-channel for electron-positron production. It is easy to see that performing the shift of variables $L_1 \leftrightarrow L_2$ produces the $u$-channel. 
We write the result:
\bea
\label{bfinal}
&&(1-n_f(p^0))\cdot{\rm Re}\,\big[\,\hat{M}^{b}\,\big] \\
&&=  
\sum_{perms}\int dL_1\int dL_2 \;\delta_{P+P_2-L_1-L_2}\;{\cal N}_b \;\Big(m^{t\dagger}_{e^+ e^-\rightarrow \gamma\gamma}\cdot m^{t}_{e^+ e^-\rightarrow \gamma\gamma} + m^{u\dagger}_{e^+ e^-\rightarrow \gamma\gamma}\cdot m^{u}_{e^+ e^-\rightarrow \gamma\gamma}\Big)~\Delta(L_1)\;\Delta(L_2)\nonumber
\eea

\subsubsection{Diagram (f)}
\label{f}

The keldysh structure of diagram (f) is identical to that of diagram (b). In this case the fermion lines are cut instead of the photon lines. 
The diagram and the amplitudes that result from taking the real part are shown in Fig. \ref{diagF}.
\par\begin{figure}[H]
\begin{center}
\includegraphics[width=8cm]{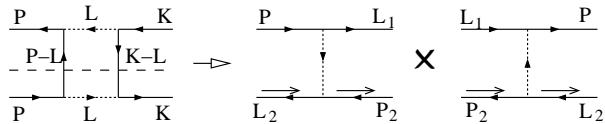}
\end{center}
\caption{The real part $\hat M^{(f)}$  produces $|m^t_{e^+ e^-\rightarrow e^+ e^-}|^2$ }
 \label{diagF}
\end{figure}
Using the first line in (\ref{scat-vars}) we write the result:
\bea
&&(1-n_f(p^0))\cdot {\rm Re}\,\big[\,\hat{M}^{f}\,\big] \\
&&~~~~= 2 \sum_{perms}\int dL_1\int dL_2\;\delta_{P+P_2-L_1-L_2}\;{\cal N}_f \;({\bf m}_f^\dagger)^{ss'\rightarrow s_2 s_3}\cdot ({\bf n}_f)^{ss'\rightarrow s_2 s_3}~\Delta(L_1)\;\Delta(L_2)\nonumber\\[2mm]
&&~~~~ ({\bf m}^\dagger_f)^{ss'\rightarrow s_2 s_3}=({\bf n}^\dagger_f)^{ss'\rightarrow s_2 s_3} = e^2\,\big(\bar u^s(P)\,\gamma^\mu\,u^{s_2}(L_1)\big)\;
 D_{\mu\mu'}^{ret}(P-L_1)\;\big(\bar v^{s_3}(L_2)\,\gamma^{\mu'}\,v^{s'}(P_2)\big)\nonumber
\eea
The amplitude $({\bf m}_f)^{ss'\rightarrow s_2 s_3}$ corresponds to the $t$-channel for electron-positron scattering. The result is:
\bea
\label{ffinal}
&&(1-n_f(p^0))\cdot{\rm Re}\,\big[\,\hat{M}^{f} \,\big]\\
&&=  2 \sum_{perms}\int dL_1\int dL_2\;\delta_{P+P_2-L_1-L_2}\;{\cal N}_f  \;\Big(m^{t\dagger}_{e^+ e^-\rightarrow e^+ e^-}\cdot m^{t}_{e^+ e^-\rightarrow e^+ e^-}\Big)~\Delta(L_1)\;\Delta(L_2)\nonumber
\eea

\subsubsection{Diagram (d)}
\label{d}

Diagram (d) and its real part are shown in Fig. \ref{diagD}. 
\par\begin{figure}[H]
\begin{center}
\includegraphics[width=10cm]{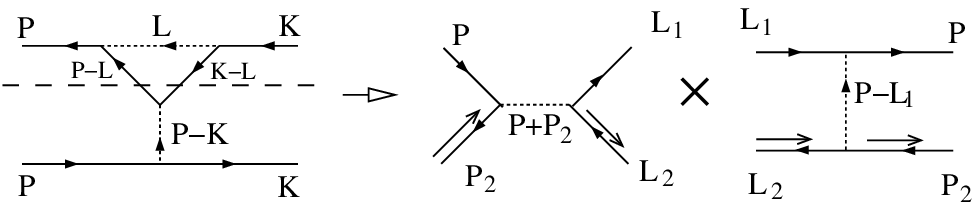}
\end{center}
\caption{The real part of $M^{(d)}$ gives $m^{t\dagger}_{e^+ e^-\rightarrow e^+ e^-}\cdot m^{s}_{e^+ e^-\rightarrow e^+ e^-}$}
 \label{diagD}
\end{figure}
Using the first line in (\ref{scat-vars}) we write the result:
\bea
&&(1-n_f(p^0))\cdot{\rm Re}\,\big[\,\hat{M}^{d}\,\big] \\
&&~~~~=-2\sum_{perms} \int dL_1\int dL_2\;\delta_{P+P_2-L_1-L_2}\;{\cal N}_f \;({\bf m}_d^\dagger)^{ss'\rightarrow s_2 s_3}\cdot ({\bf n}_d)^{ss'\rightarrow s_2 s_3}~\Delta(L_1)\;\Delta(L_2)\nonumber\\
&&~~~~ ({\bf m}^\dagger_d)^{ss'\rightarrow s_2 s_3}= e^2\,\big(\bar u^s(P)\,\gamma^\mu\;u^{s_2}(L_1)\big)
\;D^{ret}_{\mu\mu'}(P-L_1)\;\big(\bar v^{s_3}(L_2)\,\gamma^{\mu'}\,v^{s'}(P_2)\big)\nonumber\\
&&~~~~({\bf n}_d)^{ss'\rightarrow s_2 s_3} =e^2\,\big(\bar u^{s_2}(L_1)\,\gamma^\nu\,v^{s_3}(L_2)\big)\; D^{prin}_{\nu\nu'}(P+P_2)\;\big(\bar v^{s'}(P_2)\,\gamma^{\nu'}\,u^{s}(P)\big)\nonumber
\eea
The amplitude $({\bf m}_d)^{ss'\rightarrow s_2 s_3}$ corresponds to the $t$-channel for electron-positron scattering and the amplitude $({\bf n}_d)^{ss'\rightarrow s_2 s_3}$  is the $s$-channel.
The result is: 
\bea
\label{dfinal}
&&(1-n_f(p^0))\cdot{\rm Re}\,\big[\,\hat{M}^{d}\,\big] \\
&&=  -2 \sum_{perms}\int dL_1\int dL_2\;\delta_{P+P_2-L_1-L_2}\;{\cal N}_f \;\Big(m^{t\dagger}_{e^+ e^-\rightarrow e^+ e^-}\cdot m^{s}_{e^+ e^-\rightarrow e^+ e^-}\Big)~\Delta(L_1)\;\Delta(L_2)\nonumber
\eea

\subsubsection{Diagram (e)}
\label{e}

Diagram (e) and its real part are shown in Fig. \ref{diagE}. 
\par\begin{figure}[H]
\begin{center}
\includegraphics[width=10cm]{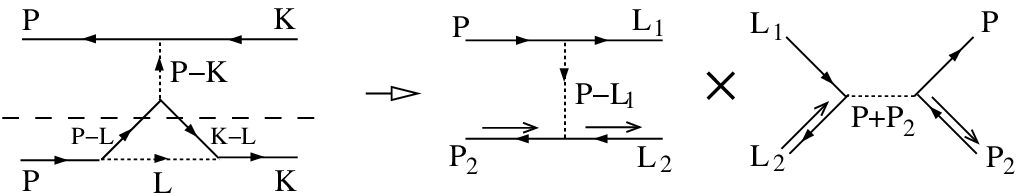}
\end{center}
\caption{The real part of $M^{(e)}$ gives $m^{s\dagger}_{e^+ e^-\rightarrow e^+ e^-}\cdot m^{t}_{e^+ e^-\rightarrow e^+ e^-}$}
 \label{diagE}
\end{figure}
It is easy to show that 
$
{\rm Re}\,\big[\,\hat{M}^e\,\big]= \big({\rm Re}\,\big[\,\hat{M}^d\,\big]\big)^\dagger
$
and therefore
\bea
\label{efinal}
&&(1-n_f(p^0))\cdot{\rm Re}\,\big[\,\hat{M}^{e}\,\big]\\
&& =  -2 \sum_{perms}\int dL_1\int dL_2\;\delta_{P+P_2-L_1-L_2}\;{\cal N}_f \; \;\Big(m^{s\dagger}_{e^+ e^-\rightarrow e^+ e^-}\cdot m^{t}_{e^+ e^-\rightarrow e^+ e^-}\Big)~\Delta(L_1)\;\Delta(L_2)\nonumber
\eea

\subsubsection{Diagram (g)}
\label{g}

Diagram (g) and its real part are shown in  Fig. \ref{diagG}.
\par\begin{figure}[H]
\begin{center}
\includegraphics[width=8cm]{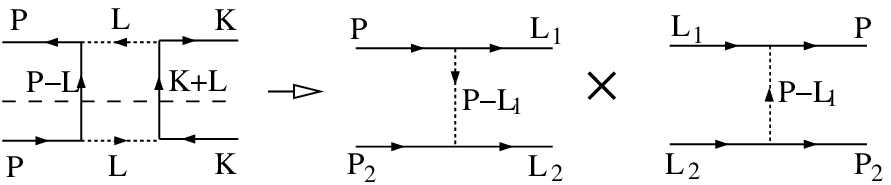}
\end{center}
\caption{The real part of $M^{(g)}$ gives $|m^t_{e^- e^-\rightarrow e^- e^-}|^2$}
 \label{diagG}
\end{figure}
Using the first line in (\ref{scat-vars2}) we obtain
\bea
&&(1-n_f(p^0))\cdot{\rm Re}\,\big[\,\hat{M}^{g}\,\big] \\
&&~~~~= 2 \sum_{perms}\int dL_1\int dL_2\;\delta_{P+P_2-L_1-L_2}\;{\cal N}_f   \;({\bf m}^\dagger_g)^{ss'\rightarrow s_2 s_3}\cdot ({\bf n}_g)^{ss'\rightarrow s_2 s_3}~\Delta(L_1)\;\Delta(L_2)\nonumber\\
&&~~~~ ({\bf m}^\dagger_g)^{ss'\rightarrow s_2 s_3}=({\bf n}^\dagger_g)^{ss'\rightarrow s_2 s_3}= e^2\,\big(\bar u^s(P)\,\gamma_\mu\,u^{s_2}(L_1)\big)\;D_{\mu\mu'}^{ret}(P-L_1)\;\big(\bar u^{s'}(P_2)\,\gamma_{\mu'}\,u^{s_3}(L_2)\big)
\nonumber
\eea
The amplitude $({\bf m}_g)^{ss'\rightarrow s_2 s_3}$ corresponds to the $t$-channel for electron-electron production. It is easy to see that performing the shift of variables $L_1\leftrightarrow L_2$ produces the $u$-channel. We write the result:
\bea
\label{gfinal}
&&(1-n_f(p^0))\cdot{\rm Re}\,\big[\,\hat{M}^{g}\,\big] \\
&&=  \sum_{perms}\int dL_1\int dL_2\;\delta_{P+P_2-L_1-L_2}\;{\cal N}_f  \;\Big(m^{t\dagger}_{e^- e^-\rightarrow e^- e^-}\cdot m^{t}_{e^- e^-\rightarrow e^- e^-}+ m^{u\dagger}_{e^- e^-\rightarrow e^- e^-}\cdot m^{u}_{e^- e^-\rightarrow e^- e^-}\Big)~\Delta(L_1)\;\Delta(L_2)\nonumber
\eea

\subsubsection{Diagram (a)}
\label{a}

Now we consider diagram (a) in Fig. \ref{bSw}. We obtain the binary contribution to the conductivity from the diagram with a one loop insertion on the vertical line. This contribution and its real part are shown in  Fig. \ref{diagA}.
\par\begin{figure}[H]
\begin{center}
\includegraphics[width=10cm]{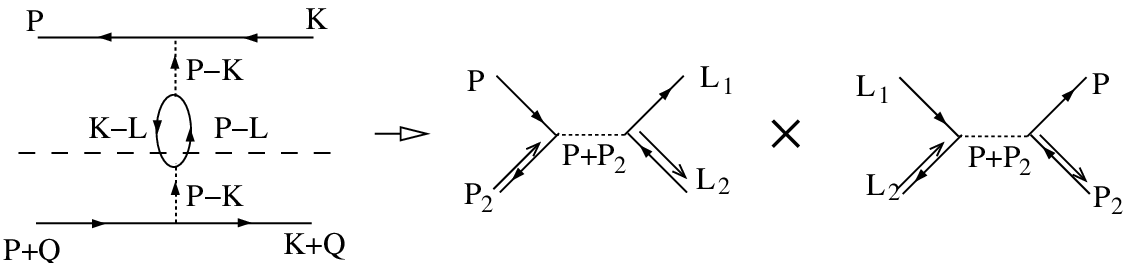}
\end{center}
\caption{The real part of $M^{(a)}$ gives  $|m^{s}_{e^+ e^-\rightarrow e^+ e^-}|^2$}
 \label{diagA}
\end{figure}
Using the first line in (\ref{scat-vars}) we write the result:
\bea
&&(1-n_f(p^0))\cdot{\rm Re}\,\big[\,\hat{M}^{a}\,\big] \\
&&~~~~= 2 \sum_{perms} \int dL_1 \int dL_2\;\delta_{P+P_2-L_1-L_2}\;{\cal N}_f \;({\bf m}^\dagger_a)^{ss'\rightarrow s_2 s_3}\cdot ({\bf n}_a)^{ss'\rightarrow s_2 s_3}~\Delta(L_1)\;\Delta(L_2)\nonumber\\
&&~~~~ ({\bf m}^\dagger_a)^{ss'\rightarrow s_2 s_3}=({\bf n}^\dagger_a)^{ss'\rightarrow s_2 s_3}= e^2\,\big(\bar u^s(P)\;
\gamma^\mu\;v^{s'}(P_2)\big)\;D^{prin}_{\mu\mu'}(P+P_2)\;\big(\bar v^{s_3}(L_2)\;\gamma^{\mu'}
\;u^{s_2}(L_1)\big)
\nonumber
\eea
The amplitude $({\bf m}_a)^{ss'\rightarrow s_2 s_3}$ corresponds to the $s$-channel for electron-positron scattering. We obtain:
\bea
\label{afinal}
&&(1-n_f(p^0))\cdot{\rm Re}\,\big[\,\hat{M}^{a}\,\big]\\
&&~~~~ =  2 \sum_{perms}\int dL_1\int dL_2\;\delta_{P+P_2-L_1-L_2}\;\;{\cal N}_f  \;\Big(m^{s\dagger}_{e^+ e^-\rightarrow e^+ e^-}\cdot m^{s}_{e^+ e^-\rightarrow e^+ e^-}\Big)~\Delta(L_1)\;\Delta(L_2)\nonumber
\eea

\subsubsection{Diagram (c)}
\label{c}

For diagram (c) is the most difficult to handle because there are two non-zero cuts. These cuts are shown in Figs. \ref{diagCh} and \ref{diagCv}. We will call them the horizontal and vertical cuts. 
\par\begin{figure}[H]
\begin{center}
\includegraphics[width=10cm]{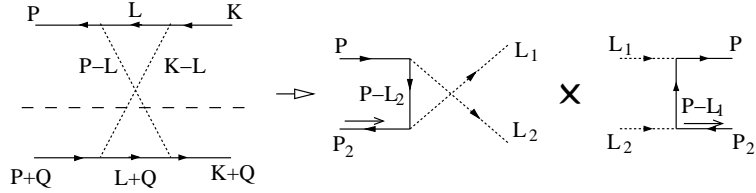}
\end{center}
\caption{The horizontal cut of the real part of $M^{(c)}$ gives $m^{t\dagger}_{e^- e^+\rightarrow \gamma \gamma}\cdot m^{u}_{e^- e^+\rightarrow \gamma\gamma}+m^{u\dagger}_{e^- e^+\rightarrow \gamma \gamma}\cdot m^{t}_{e^- e^+\rightarrow \gamma\gamma}$}
 \label{diagCh}
\end{figure}
\par\begin{figure}[H]
\begin{center}
\includegraphics[width=10cm]{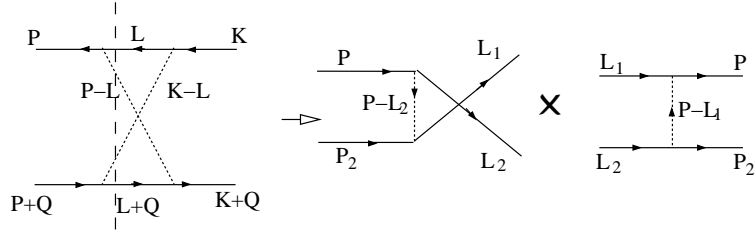}
\end{center}
\caption{The vertical cut of the real part of $M^{(c)}$ gives $m^{t\dagger}_{e^- e^-\rightarrow e^- e^-}\cdot m^{u}_{e^- e^-\rightarrow e^- e^-}+
m^{u\dagger}_{e^- e^-\rightarrow e^- e^-}\cdot m^{t}_{e^- e^-\rightarrow e^- e^-}$}
 \label{diagCv}
\end{figure}
For the horizontal cut we use the first line in (\ref{scat-vars}). For the vertical cut we shift variables $L\to -L+P$ and use the first line in (\ref{scat-vars2}). We obtain:
\bea
&&(1-n_f(p^0))\cdot{\rm Re}\,\big[\,\hat{M}^{c-horz}\,\big] \\
&&= \sum_{perms}\int dL_1\int dL_2\;\delta_{P+P_2-L_1-L_2}\;{\cal N}_b\;\Delta(L_1)\;\Delta(L_2)\bigg({\bf m}_{c-horz}^{\dagger~ss'\rightarrow s_2 s_3}\cdot {\bf n}_{c-horz}^{ss'\rightarrow s_2 s_3}+(L_1\leftrightarrow L_2)\bigg)\nonumber\\
&& (1-n_f(p^0))\cdot{\rm Re}\,\big[\, \hat{M}^{c-vert}\,\big] \nonumber\\
&&= -\sum_{perms}\int dL_1\int dL_2\;\delta_{P+P_2-L_1-L_2}\;{\cal N}_f\;\Delta(L_1)\;\Delta(L_2)\Big({\bf m}_{c-vert}^{\dagger~ss'\rightarrow s_2 s_3}\cdot {\bf n}_{c-vert}^{ss'\rightarrow s_2 s_3}~+~(L_1\leftrightarrow L_2)\Big)\nonumber\\[4mm]
&& ({\bf m}^\dagger_{c-horz})^{ss'\rightarrow \lambda \lambda'}=e^2\, \bar u^s_\alpha(P)
\big(\gamma^\mu\;S^{ret}(P-L_1)\;\gamma^\nu\big)_{\alpha\delta}\;v_\delta(P_2)\;\epsilon^\lambda_\mu(L_1)\;\epsilon^{\lambda'}_\nu(L_2)\nonumber\\
&& ({\bf n}_{c-horz})^{ss'\rightarrow \lambda \lambda'}= e^2\,\bar v(P_2)_\gamma\;\big(\gamma^{\mu'}S^{adv}(P-L_2)\gamma^{\nu'}\big)_{\gamma\beta}\;u(P)_\beta\;\epsilon^{\lambda *}_{\mu'}(L_1)\;\epsilon^{\lambda '*}_{\nu'}(L_2)\nonumber\\
&&({\bf m}^{\dagger}_{c-vert})^{ss'\rightarrow \lambda \lambda'}=e^2\,
\big(\bar u(P)\;\gamma^\mu\;u(L_1)\big)\;D^{ret}_{\mu\mu'}(P-L_1)\;\big(\bar u(P_2)\;\gamma^{\mu'}\;u(L_2)\big)\nonumber\\
&&({\bf n}_{c-vert})^{ss'\rightarrow \lambda \lambda'}=e^2\,\big(\bar u(L_2)\;\gamma^\nu\;u(P)\big)\;D^{adv}_{\nu\nu'}(P-L_2)\;\big(\bar u(L_1)\;\gamma^\nu\;u(P_2)\big)\nonumber
\eea
 We write the results:
\bea
\label{chfinal}
&& (1-n_f(p^0))\cdot{\rm Re}\,\big[\,\hat{M}^{c-horz}\,\big] \\
&&=  \sum_{perms}\int dL_1\int dL_2\;\delta_{P+P_2-L_1-L_2}\;{\cal N}_b\;\Delta(L_1)\;\Delta(L_2)\;\Big(m^{t\dagger}_{e^- e^+\rightarrow \gamma \gamma}\cdot m^{u}_{e^- e^+\rightarrow \gamma\gamma}~~+~~m^{u\dagger}_{e^- e^+\rightarrow \gamma \gamma}\cdot m^{t}_{e^- e^+\rightarrow \gamma\gamma}\Big)\nonumber\\[2mm]
\label{cvfinal}
&& (1-n_f(p^0))\cdot{\rm Re}\,\big[\,\hat{M}^{c-vert}\,\big] \\
&&= -\sum_{perms}\int dL_1\int dL_2\;\delta_{P+P_2-L_1-L_2}\;{\cal N}_f\;\Delta(L_1)\;\Delta(L_2)\;\Big(m^{t\dagger}_{e^- e^-\rightarrow e^- e^-}\cdot m^{u}_{e^- e^-\rightarrow e^- e^-}~~+~~
m^{u\dagger}_{e^- e^-\rightarrow e^- e^-}\cdot m^{t}_{e^- e^-\rightarrow e^- e^-}\Big)\nonumber
\eea

\subsubsection{Combine Results}

Including the contributions from all diagrams we have 
\bea
\label{bigM}
\sum_{j\in \{a,b,c,d,e,f,g\}}{\rm Re}\,\big[\,\hat M^{(j)}(P,K) \,\big] = {\rm Re}\,\big[\,\hat {M}_{e^+e^-\to \gamma\gamma}\,\big]+{\rm Re}\,\big[\,\hat {M}_{e^-e^-\to e^-e^-} \,\big]+{\rm Re}\,\big[\,\hat {M}_{e^+e^-\to e^+e^-}\,\big]
\eea
Combining (\ref{bfinal}), (\ref{chfinal}) we obtain:
\bea
\label{gammaf}
&&(1-n_f(p^0))\cdot{\rm Re}\,\big[\,\hat {M}_{e^+e^-\to \gamma\gamma}\,\big] \\
= &&\sum_{perms}\int dL_1\int dL_2\;\delta_{P+P_2-L_1-L_2}\;{\cal N}_b\;\Delta(L_1)\;\Delta(L_2)\big|m^t_{e^+e^-\to \gamma\gamma}+m^u_{e^+e^-\to \gamma\gamma}\big|^2\nonumber
\eea
Combining (\ref{gfinal}), (\ref{cvfinal}) we obtain:
\bea
\label{elecf}
&&(1-n_f(p^0))\cdot{\rm Re}\,\big[\,\hat {M}_{e^-e^-\to e^-e^-} \,\big]\\
&&=\sum_{perms} \int dL_1\int dL_2\;\delta_{P+P_2-L_1-L_2}\;{\cal N}_f\;\Delta(L_1)\;\Delta(L_2)\big|m^t_{e^-e^-\to e^-e^-}-m^u_{e^-e^-\to e^-e^-}\big|^2\nonumber
\eea
Combining (\ref{ffinal}), (\ref{dfinal}), (\ref{efinal}), (\ref{afinal}) we obtain:
\bea
\label{posf}
&&(1-n_f(p^0))\cdot{\rm Re}\,\big[\,\hat {M}_{e^+e^-\to e^+e^-}\,\big] \\
&&=  2\sum_{perms}\int dL_1\int dL_2\;\delta_{P+P_2-L_1-L_2}\;{\cal N}_f\;\Delta(L_1)\;\Delta(L_2)\big|m^t_{e^+e^-\to e^+e^-}-m^s_{e^+e^-\to e^+e^-}\big|^2\nonumber
\eea

\subsection{Integral Equation}
\label{IntEqn}

Finally,  we substitute (\ref{bigM}), (\ref{gammaf}), (\ref{elecf}) and (\ref{posf}) into the integral equation (\ref{AMY0}).
We note that $\rho(-P_2)B^i(-P_2) = \rho(P_2)B^i(P_2)$ so that both of the definitions $P_2=K$ and $P_2=-K$ produce the same overall factor. 
Combining all terms the integral equation (\ref{AMY0}) becomes:
\bea
\label{bSv}
&&2 (1-n_f(p^0))\cdot{\rm Im}\,\hat \Sigma(P) \cdot B^i(P) \\[2mm]
&&~~~~= (1-n_f(p^0))\cdot{\rm Re}\,\hat\Lambda_0^i(3,P) \nonumber\\[2mm]
&&~~~~+\frac{1}{2}\int dP_2\int dL_1\int dL_2\;\delta_{P+P_2-L_1-L_2}\Delta(P_2)\Delta(L_1)\Delta(L_2)\sum_{perms}\bigg[\;\big|M\big|^2 \;B^i(P_2)\bigg]\nonumber
\eea
where 
\bea
\label{me1}
&&\big|M\big|^2 \\
&&= {\cal N}_b\;\big|m^t_{e^+e^-\to \gamma\gamma}+m^u_{e^+e^-\to \gamma\gamma}\big|^2+{\cal N}_f\;\big|m^t_{e^-e^-\to e^-e^-}-m^u_{e^-e^-\to e^-e^-}\big|^2+2{\cal N}_f\;\big|m^t_{e^+e^-\to e^+e^-}-m^s_{e^+e^-\to e^+e^-}\big|^2\nonumber
\eea
The next step is to show that the factor $\big|M\big|^2$ can be written in the form 
\bea
\label{me2}
\big|M\big|^2 = \sum_{perms}\bigg[{\cal N}_f\;\big|m^t_{e^-e^-\to e^-e^-}-m^u_{e^-e^-\to e^-e^-}\big|^2
+{\cal N}_b\;\big|m^t_{e^+e^-\to \gamma\gamma}+m^u_{e^+e^-\to \gamma\gamma}\big|^2\bigg]
\eea
There are two points to address:

\xx (1) The last term in (\ref{me1}) can be written in a more symmetric way as
\bea
2{\cal N}_f\;\big|m^t_{e^+e^-\to e^+e^-}-m^s_{e^+e^-\to e^+e^-}\big|^2 \to {\cal N}_f\;\bigg[\big|m^t_{e^+e^-\to e^+e^-}-m^s_{e^+e^-\to e^+e^-}\big|^2
+\big|m^u_{e^+e^-\to e^+e^-}-m^s_{e^+e^-\to e^+e^-}\big|^2\bigg]
\eea
Using (\ref{defn-perms}) and this result the last two terms in (\ref{me1}) can be written:
\bea
\label{fermstuff}
\sum_{perms} {\cal N}_f\;\big|m^t_{e^-e^-\to e^-e^-}-m^u_{e^-e^-\to e^-e^-}\big|^2
\eea
The three terms contained in the sum in (\ref{fermstuff}) correspond to the familiar results for bhabha and m\o ller scattering, and are shown in Fig. \ref{fermscat}. 
\par\begin{figure}[H]
\begin{center}
\includegraphics[width=12cm]{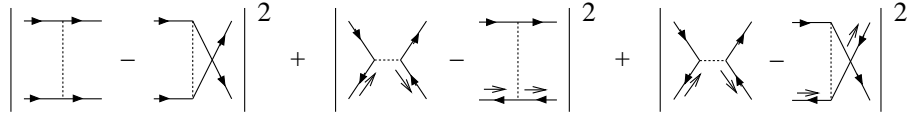}
\end{center}
\caption{Bhabha and M\o ller scattering}
 \label{fermscat}
\end{figure}

\xx (2) The first term in (\ref{me1}) is the square of the amplitude that corresponds to electron-positron production. The diagrams are shown in Fig. \ref{gammascat}. 
\par\begin{figure}[H]
\begin{center}
\includegraphics[width=3cm]{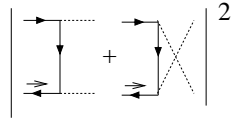}
\end{center}
\caption{electron-positron production}
 \label{gammascat}
\end{figure}
The terms that give compton scattering are missing because they don't contribute to the conductivity (since they do not connect with the two pinched pairs of fermion propagators). In order to write the matrix element in an symmetric way we must temporarily include these compton terms:
\bea
n_b(p^0)(1-n_f(l^0_1))(1-n_b(l^0_2))\big|m^s_{e^+e^-\to \gamma\gamma}+m^u_{e^+e^-\to \gamma\gamma}\big|^2\\
n_b(p^0)(1-n_b(l^0_1))(1-n_f(l^0_2))\big|m^t_{e^+e^-\to \gamma\gamma}+m^s_{e^+e^-\to \gamma\gamma}\big|^2\nonumber
\eea
which allows us to write $\big|M\big|^2$ in the form given in Eqn. (\ref{me2}).

Next we substitute (\ref{me2}) into (\ref{bSv}) and use:
\bea
\sum_{perms}\bigg[f(P,P_2,L_1,L_2)\cdot \sum_{perms}f'(P,P_2,L_1,L_2)\bigg] = \sum_{perms}f(P,P_2,L_1,L_2)\cdot \sum_{perms}f'(P,P_2,L_1,L_2)
\eea
to obtain:
\bea
\label{bSu}
&&2(1-n_f(p^0))\cdot {\rm Im}\,\hat \Sigma(P)\cdot B^i(P) \\
&&~~~~= (1-n_f(p^0))\cdot{\rm Re}\,\hat\Lambda_0^i(3,P) +\frac{1}{2}\int dP_2\int dL_1\int dL_2\;\delta_{P+P_2-L_1-L_2}\; \Delta(P_2)\Delta(L_1)\Delta(L_2)\; \big|M\big|^2\;\sum_{perms} \;B^i(P_2) \nonumber\\
&&~~~~= (1-n_f(p^0))\cdot{\rm Re}\,\hat\Lambda_0^i(3,P) \nonumber\\[2mm]
&&~~~~+\frac{1}{2}\int dP_2\int dL_1\int dL_2\;\big|M\big|^2\;\delta_{P+P_2-L_1-L_2}\Delta(P_2)\Delta(L_1)\Delta(L_2)\; \;[B^i(P_2)-B^i(L_1)-B^i(L_2)] \nonumber
\eea
where we have used $\Delta(-X)=\Delta(X)$ and $B^i(-X)=-B^i(X)$.\\

The last step is to rearrange (\ref{bSu}) in the form:
\bea
\label{AMY1}
&&-(1-n_f(p^0))\cdot{\rm Re}\,\hat\Lambda_0^i(3,P) \\
&&~~~~= -2 (1-n_f(p^0))\cdot{\rm Im}\,\hat \Sigma(P)\cdot B^i(P)\nonumber\\[2mm]
&&~~~~+\frac{1}{2}\int dP_2\int dL_1\int dL_2\;\big|M\big|^2\;\delta_{P+P_2-L_1-L_2} \Delta(P_2)\Delta(L_1)\Delta(L_2)\; \;[B^i(P_2)-B^i(L_1)-B^i(L_2)] \nonumber
\eea
and obtain an expression for ${\rm Im}\hat\Sigma(P)$ in terms of the squared matrix element $\big|M\big|^2$.  
Using (\ref{dyson}) and the terms shown in Fig. \ref{Phi} we have two contributions to $\Sigma(P)$ which are shown in Fig. \ref{sigma1}. 
\par\begin{figure}[H]
\begin{center}
\includegraphics[width=8cm]{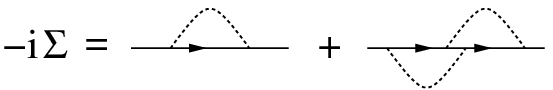}
\end{center}
\caption{Contributions to $\Sigma(P)$ from $\Phi[S,D]$}
 \label{sigma1}
\end{figure}
We expand the propagators in the first diagram and extract the terms that correspond to one-loop insertions on each line. Combining, we obtain the three diagrams shown in Fig. \ref{2nd0-new2}.
\par\begin{figure}[H]
\begin{center}
\includegraphics[width=10cm]{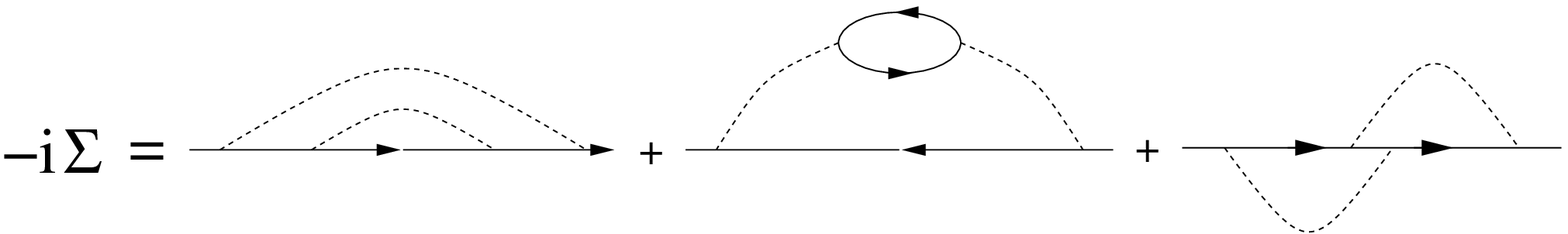}
\end{center}
\caption{Expanded contributions to $\Sigma(P)$}
 \label{2nd0-new2}
\end{figure}
These three graphs can be obtained from a tadpole graph of the form shown in Fig. \ref{tadpole}.
\par\begin{figure}[H]
\begin{center}
\includegraphics[width=3cm]{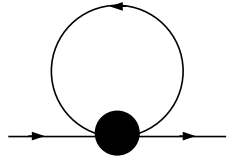}
\end{center}
\caption{Representation of $\Sigma(P)$ in terms of a tadpole diagram}
 \label{tadpole}
\end{figure}
We do not need to include all of the terms $\hat M^{(j)}$ shown in Fig. \ref{bSx}. Using $\hat M^{(b)}$ in Fig. \ref{tadpole} produces the first graph in Fig. \ref{2nd0-new2}; using $\hat M^{(c)}$ produces the second graph; and $\hat M^{(g)}$ produces the third graph. Summing over keldysh indices using the program in \cite{MCTF} we obtain:
\bea
\Sigma^{ret}_{\alpha\beta}(P) &=& \frac{i}{2}\sum_{j\in\{b,c,g\}}\int dK\\ 
&&\cdot \big(M^{(j)}_{\alpha\beta;\gamma\delta}(11,P,K)S^{ret}_{\delta\gamma}(K) +M^{(j)}_{\alpha\beta;\gamma\delta}(7,P,K)S^{adv}_{\delta\gamma}(K)+M^{(j)}_{\alpha\beta;\gamma\delta}(3,P,K)N_F(K)S^{d}_{\delta\gamma}(K)\big)\nonumber
\eea
We multiply by $\Psl$ and take the trace. We write the fermion propagator as $S(K)=\Ksl G(K)$ (see Eqn. (\ref{S-pinch})) and extract $\hat M$ using (\ref{Mhat}). Taking the imaginary part we obtain:
\bea
{\rm Im}\hat\Sigma(P) = -\frac{1}{2}\sum_{j\in\{b,c,g\}}\int dK {\rm Re}\,\big[\,\hat M^{(j)}(7,P,K)-\hat M^{(j)}(11,P,K)-N_F(K)\hat M^{(j)}(3,P,K)\,\big]\,\rho(K)
\eea
Using the same method as before it is straightforward to show that the result is:
\bea
(1-n_f(p^0))\cdot{\rm Im}\hat\Sigma(P) &&=-\frac{1}{4}\int dP_2\int dL_1\int dL_2\;\delta_{P+P_2-L_1-L_2}\\
&&\cdot\sum_{perms}\bigg[{\cal N}_f\;\big|m^t_{e^-e^-\to e^-e^-}-m^u_{e^-e^-\to e^-e^-}\big|^2
+{\cal N}_b\;\big|m^t_{e^+e^-\to \gamma\gamma}+m^u_{e^+e^-\to \gamma\gamma}\big|^2\bigg]\;\Delta(P_2)\Delta(L_1)\Delta(L_2)\nonumber\\
&&=-\frac{1}{2}\int dP_2\int dL_1\int dL_2\;\delta_{P+P_2-L_1-L_2}\;\big|M\big|^2\;\Delta(P_2)\Delta(L_1)\Delta(L_2)\nonumber
\eea
where we have used (\ref{me2}) in the last line. 
Substituting into (\ref{AMY1}) gives:
\bea
\label{AMY2}
&&-(1-n_f(p^0))\cdot{\rm Re}\,\hat\Lambda_0^i(3,P) \\
&&= \frac{1}{2}\int dP_2\int dL_1\int dL_2\;\big|M\big|^2\;\delta_{P+P_2-L_1-L_2}\Delta(P_2)\Delta(L_1)\Delta(L_2)\; \;[B^i(P)+B^i(P_2)-B^i(L_1)-B^i(L_2)] \nonumber
\eea
This equation is exactly the same as that obtained in \cite{AMY} using kinetic theory. 

\section{Conclusions}
\label{Conclusions}

In this paper we have studied the applicability of the 2PI effective action to describe the equilibration of quantum fields.
We have calculated the QED electrical conductivity using the 2PI effective action. We have used a modified version of the usual 2PI effective action which is defined with respect to self-consistent solutions of the 2-point functions. The green functions obtained from this modified effective action satisfy ward identities and  the conductivity obtained from the kubo relation is explicitly gauge invariant. We have shown that using this method the summation over ladder graphs is obtained automatically, without any power counting analysis. 

We have also done an explicit calculation at 3-loop order. We have calculated the integral equation that determines the conductivity and shown that the full matrix element corresponding to all binary scattering and production processes is obtained. 
Our result is complete at leading-log order but it does not contain all contributions at leading order since the colinear terms are not included. These terms will be present in a calculation using the 3-loop 3PI effective action, and this work is currently in progress.
Furthermore, the method we have developed in this paper should be generalizable to the calculation of other transport coefficients.
 
Our calculation provides a field theoretic connection to the kinetic theory results of \cite{AMY}, which is useful in itself. In addition, it seems likely that quantum field theory provides a better framework than kinetic theory for calculations beyond leading order. Our results provide strong support for the use of $n$PI effective theories as a method to study the equilibration of quantum fields.

\appendix 

\section{}

\label{appendixA}

In this appendix, we give some details of the method we use to calculate ${\rm Re}\,\big[\,\hat M^{(j)}(P,K)\,\big]$ for each of the diagrams in Fig. \ref{bSz}. The basic strategy is explained briefly in section \ref{ME}.  Throughout this appendix, we simplify the notation by considering scalar fields with a cubic iteraction ($\phi^3$ theory). The dirac and lorentz structure of the fermion and photon fields will not effect the basic form of the calculation. Note that for $\phi^3$ theory (\ref{Mhat}) becomes ${\rm Re}\,\big[\,\hat M^{(j)}(P,K)\,\big] \Rightarrow -i\, {\rm Im}\,\big[\,{\bf M}^{(j)}(P,K)\,\big]$. For simplicity we set the coupling constant to one. \\

\xx There are four basic steps to the calculation:\\

\xx Step (1): We use the mathematica program in \cite{MCTF} to perform the sum over keldysh indices. This program is described in detail in  \cite{MCTF} and is available at www.brandonu.ca/physics/fugleberg/Research/Dick.html. The program can be used to calculate the integrand corresponding to any diagram (up to five external legs) in the keldysh, RA or 1-2 basis. The user supplies input in the form of lists of momenta, and vertices for each propagator and vertex. \\

\xx Step (2): We extract the real part of each diagram.  The method is related to the Cutkosky rules at finite temperature, and is described in detail in \cite{MC2}. 
The basic strategy is as follows:

\xx (a) We make use of the fact that it is kinematically forbidden for three on-shell lines to meet at a vertex:
\bea
\label{kine}
d(\pm(P_1\pm P_2))d(P_1)d(P_2) = 0
\eea
\xx (b) When an integral does not contain thermal distribution functions, we can add (or subtract) terms which correspond to products of propagators with all poles on the same side of the real axis, since these terms are identically zero by contour integration.

We apply these rules to each diagram.  The basic strategy is as follows.

The real part is given by 
\bea
{\rm Re}\,\big[\,\hat M^{(j)}(P,K)\,\big]=\big[\hat M^{(j)}(P,K)+\big(\hat M^{(j)}(P,K)\big)^*\big]/2
\eea  
Each diagram has four internal propagators. Terms with an even number of cut propagators are real. There are no terms with zero cut lines, and terms with four cut lines are zero by (\ref{kine}). We are left with terms with any pair of lines cut. Some of these terms are also zero by (\ref{kine}).  

To isolate the surviving terms we must take into account the fact that the momenta $P$ and $K$ are on-shell external momenta.  As a consequence, Eqn (\ref{kine}) allows us to make the replacement 
\bea
\label{PrinPK}
G^{ret}(P-K)=G^{adv}(P-K)={\rm Prin}(P-K)
\eea
In addition we obtain relations like
\bea
\label{PrinL}
d(K-L)G^{ret}(L) ~\to ~d(K-L){\rm Prin}(L)\,;~~~~d(K-L)G^{adv}(L) ~\to ~d(K-L){\rm Prin}(L)
\eea

In diagrams where all propagators carry different momenta (for example, diagram (d)), the procedure is straightforward. In diagrams where more than one propagator carries a given momentum (for example, diagram (b)), one must be careful to show that potentially dangerous terms that contain the square of a delta function (like $\delta(L^2)$ for the case of diagram (b)) do not appear. The disappearance of these unphysical terms is a well known result due to the KMS condition \cite{lands}.\\

\xx Step (3): We relabel the momenta so that the real part of each diagram can be written as a product of amplitudes that have the usual form for scattering and production amplitudes. The method is described in section \ref{ME} and illustrated below.\\

\xx Step (4): We rewrite the thermal factors and show that they correspond to the appropriate combination of statistical emission and absorption factors. This is done using the identity (\ref{N-identity}).\\

\xx In the sections below, we apply this strategy to diagrams (b) and (d).

\subsection{Diagram (d)}

We start by looking at diagram (d) in Fig. \ref{bSz}. First we note that it is immediately clear from looking at the diagram that 
the only possible contribution comes from the horizontal cut shown in Fig. \ref{diagD}. There is no other way to cut two lines without getting zero by (\ref{kine}).

Step (1): We use the mathematica program in \cite{MCTF} to perform the sum over keldysh indices and use (\ref{PrinPK}) immediately. 
We separate the terms that depend on the thermal functions $N_B$ and $N_F$. The results are
\bea
\label{thd1}
&&{\rm Re}\,\big[\,{\hat M}_{th}^{(d)}(P,K)\,\big]=-\frac{1}{2}  \int dL\;d(K-L) \text{Prin}(P-K) \\
&&~~~~~ \big[d(P-L)
   \left(N_F\left(k_0\right)-N_F\left(k_0-l_0\right)\right) N_F\left(p_0-l_0\right)
   G^{\text{ret}}(L) \nonumber\\
&&~~~~~ +N_F\left(k_0\right) N_F\left(k_0-l_0\right) G^{\text{adv}}(P-L)
   G^{\text{ret}}(L)-N_F\left(k_0\right) N_F\left(k_0-l_0\right) G^{\text{adv}}(L)
   G^{\text{ret}}(P-L)\big]\nonumber
\eea
and 
\bea
\label{mdz}
&&{\rm Re}\,\big[\,{\hat M}_0^{(d)}(P,K)\,\big]\\
&&=\frac{1}{2}\int dL\;\big[  \text{Prin}(P-K) G^{\text{adv}}(P-L) G^{\text{ret}}(K-L) G^{\text{ret}}(L)+
    \text{Prin}(P-K) G^{\text{adv}}(K-L) G^{\text{adv}}(L) G^{\text{ret}}(P-L)\big]\nonumber
\eea

Step (2): Using (\ref{PrinL}) we can rewrite (\ref{thd1}) to obtain: 
\bea
\label{mdth}
&&{\rm Re}\,\big[\,{\hat M}_{th}^{(d)}(P,K)\,\big]\\
&&=\frac{1}{2} \int dL\; d(K-L) d(P-L) \text{Prin}(L) \text{Prin}(P-K) \left(N_F\left(k_0\right)
   N_F\left(k_0-l_0\right)-\left(N_F\left(k_0\right)-N_F\left(k_0-l_0\right)\right)
   N_F\left(p_0-l_0\right)\right)\nonumber
\eea

Next we rewrite (\ref{mdz}). We use the fact that terms that contain the triple of propagators $G^{\text{adv}}(K-L) G^{\text{adv}}(P-L) G^{\text{ret}}(L)$ or $G^{\text{ret}}(K-L) G^{\text{ret}}(P-L) G^{\text{adv}}(L)$ (and no other propagators that depend on $L$) can be added or subtracted for free, because they give zero when the integration over $l_0$ is done (since all poles lie on the same side of the real axis). Using this trick we can rewrite (\ref{mdz}) to obtain
\bea
\label{mdzz}
{\rm Re}\,\big[\,{\hat M}_0^{(d)}(P,K)\,\big]=-\frac{1}{2} \int dL\; d(K-L) d(P-L) \text{Prin}(L)
\eea
Combining (\ref{mdth}) and (\ref{mdzz}) we obtain:
\bea
{\rm Re}\,\big[\,{\hat M}^{(d)}(P,K)\,\big]&=&-\frac{1}{2} \int dL\; d(K-L) d(P-L) \text{Prin}(L)\text{Prin}(P-K)\\
&\cdot& \big[(1- N_F\left(k_0\right)
   \left(N_F\left(k_0-l_0\right)-N_F\left(p_0-l_0\right)\right)+N_F\left(k_0-l_0\right)
   N_F\left(p_0-l_0\right)\big]\nonumber
\eea

Step (3): Using (\ref{rho}) and the first line of (\ref{scat-vars}), and using the notation in (\ref{defn-perms}), we obtain:
\bea
{\rm Re}\,\big[\,{\hat M}^{(d)}\,\big]&=&-\frac{1}{2}\sum_{perms} \int dL_1\int dL_2\;\delta_{P+P_2-L_1-L_2}\; \Delta\left(L_1\right) \Delta\left(L_2\right) \\
&\cdot&\text{Prin}\left(P-L_1\right)
   \text{Prin}\left(P+P_2\right) \left(1+N_F\left(l^0_1\right)
   \left(N_F\left(l^0_2\right)-N_F\left(p^0_2\right)\right)-N_F\left(l^0_2\right)
   N_F\left(p^0_2\right)\right)\nonumber
   \eea

Step (4): Using identities of the form (\ref{N-identity}) it is easy to show that this result can be written:
\bea
(1-n_f(p^0))\cdot{\rm Re}\,\big[\,{\hat M}^{(d)}\,\big]=-2 \sum_{perms}\int dL_1\int dL_2\;\delta_{P+P_2-L_1-L_2}\;{\cal N}_f \,\Delta\left(L_1\right) \Delta\left(L_2\right) \text{Prin}(L) \text{Prin}\left(P+P_2\right)
 \eea
This equation agrees with (\ref{dfinal}), except for the dirac structure.

\subsection{Diagram (b)}

From inspection of diagram (b)  (Fig. \ref{diagB}) there are two possible pairs of propagators that could contribute to the real part: one pair corresponds to the horizontal cut and the other pair to the vertical cut. Note however that the pair of propagators involved in the vertical cut both carry the momentum $L$ and thus cutting both of these propagators would produce a divergent term of the form $\delta(L^2)^2$. We show below that the coefficient of this term is identically zero. 

Step (1): We sum over keldysh indices and use $d(L)=G^{ret}(L)-G^{adv}(L)$ to expand all factors that depend on $d(L)$. The result is:
\bea
&&{\rm Re}\,\big[\,{\hat M}^{(b)}(P,K)\,\big]=\frac{1}{2} \int dL\\
\cdot&&\bigg(\left(N_B\left(k_0-l_0\right)-N_F\left(k_0\right)\right)
   \left(N_B\left(p_0-l_0\right)+N_F\left(l_0\right)\right)d(K-L) d(P-L)  G^{\text{adv}}(L) G^{\text{ret}}(L)\nonumber\\[2mm]
&&-\left(\left[1-N_B\left(k_0-l_0\right)
   \left(N_F\left(k_0\right)-N_F\left(l_0\right)\right)-N_F\left(k_0\right)
   N_F\left(l_0\right)\right]d(K-L) -G^{\text{adv}}(K-L)\right) G^{\text{adv}}(P-L) G^{\text{ret}}(L)^2 \nonumber\\[2mm]
&&- \left(\left[1-N_B\left(k_0-l_0\right)
   \left(N_F\left(k_0\right)-N_F\left(l_0\right)\right)-N_F\left(k_0\right)
   N_F\left(l_0\right)\right]d(K-L) -G^{\text{ret}}(K-L)\right) G^{\text{ret}}(P-L)G^{\text{adv}}(L)^2\bigg)\nonumber
\eea

Step (2): Using the kms condition (\ref{N-identity}) we find that the factor in square brackets is zero. Using this result we can rewrite the terms proportional to $G^{ret}(L)^2$ and $G^{adv}(L)^2$ in the equation above as (the integrals are zero after doing the $l_0$ integration):
\bea
\frac{1}{2}\int dL\; \left(G^{\text{ret}}(K-L) G^{\text{ret}}(P-L) G^{\text{adv}}(L)^2+G^{\text{adv}}(K-L)
   G^{\text{adv}}(P-L) G^{\text{ret}}(L)^2\right) = 0
\eea
We rewrite the surviving terms using (from (\ref{N-identity}))
\bea
N_F\left(l_0\right) =  \frac{N_B\left(k_0-l_0\right)
   N_F\left(k_0\right)-1}{N_B\left(k_0-l_0\right)-N_F\left(k_0\right)}
\eea
and obtain:
\bea
&&{\rm Re}\,\big[\,{\hat M}^{(b)}(P,K)\,\big]=-\frac{1}{2} \int dL\;d(K-L) d(P-L)\\
&&~~~~ \left(1+N_B\left(p_0-l_0\right) N_F\left(k_0\right)-N_B\left(k_0-l_0\right)
   \left(N_B\left(p_0-l_0\right)+N_F\left(k_0\right)\right)\right) G^{\text{adv}}(L)
   G^{\text{ret}}(L)\nonumber
\eea

Step (3): Using the first line in (\ref{scat-vars}) and the notation in (\ref{defn-perms}) we obtain:
\bea
{\rm Re}\,\big[\,{\hat M}^{(b)}\,\big]
&=&-\frac{1}{2}\sum_{perms}\int dL_1\int dL_2\;\delta_{P+P_2-L_1-L_2}\; \Delta\left(L_1\right) \Delta\left(L_2\right)\\
&\cdot& \left(1+N_B\left(l^0_2\right)
   \left(N_B\left(l^0_1\right)-N_F\left(p^0_2\right)\right)-N_B\left(l^0_1\right)
   N_F\left(p^0_2\right)\right) G^{\text{adv}}\left(P-L_1\right) G^{\text{ret}}\left(P-L_1\right) \nonumber
   \eea

Step (4): Using identities of the form (\ref{N-identity}) it is easy to show that this result can be written:
\bea
(1-n_f(p^0))\cdot{\rm Re}\,\big[\,{\hat M}^{(b)}\,\big]&=& -2 \sum_{perms}\int dL_1\int dL_2\;{\cal N}_b\; \Delta\left(L_1\right) \Delta\left(L_2\right) G^{\text{adv}}\left(P-L_1\right)
   G^{\text{ret}}\left(P-L_1\right)\nonumber
\eea

We compare this result with (\ref{bfinal}). There is an extra minus sign in (\ref{bfinal}) which comes from the sign associated with the factor $v(P_2)\bar v(P_2)$ (see Eqn (\ref{cut-lines})).\\

\large

\xx {\bf Acknowledgements:}

\normalsize

\xx The authors thank Gert Aarts for useful discussions. 

\xx This work was supported by the Natural Sciences and Engineering Research Council of Canada.

\newpage

\end{document}